\documentclass[12pt,a4paper]{article}
\pdfoutput=1

\usepackage{geometry}
\usepackage[numbers,sort&compress]{natbib}
\usepackage{amsmath}
\usepackage{amssymb}
\usepackage[dvips]{graphicx}
\usepackage{pstricks}
\usepackage{bm}
\usepackage{pbox}
\usepackage{placeins}
\usepackage{graphicx}
\usepackage{caption}
\usepackage{subcaption}
\usepackage[T1]{fontenc}
\usepackage{footnote}
\usepackage{pdfpages}
\usepackage{hhline}
\usepackage{multirow}
\usepackage{enumitem}
\usepackage[nottoc,notlot,notlof]{tocbibind}
\usepackage[title,titletoc]{appendix}
\usepackage{dsfont}
\allowdisplaybreaks

\usepackage{cases}

\renewcommand{\thetable}{\Roman{table}}
\newcommand\scalemath[2]{\scalebox{#1}{\mbox{\ensuremath{\displaystyle #2}}}}

\definecolor{MyDarkBlue}{rgb}{0.1, 0.1, 0.8} 
\definecolor{MyLightBlue}{rgb}{0.22,0.51,0.9}
\definecolor{MyGreen}{rgb}{0.0, 0.5, 0.0}
\definecolor{BrickRed}{rgb}{0.8, 0.25, 0.33}
\usepackage[colorlinks=true,linkcolor=blue,citecolor=MyDarkBlue,
urlcolor=MyLightBlue,bookmarksnumbered=true,bookmarksopen]{hyperref}
\hypersetup{colorlinks, citecolor=blue,linkcolor=red, urlcolor=MyLightBlue}

\begin{document}
\vspace*{-0.2in}
\begin{flushright}
OSU-HEP-20-03
\end{flushright}
\vspace{0.5cm}
\begin{center}
{\Large\bf 
A Common Origin of Neutrino Masses and\\  $R_{D^{(\ast)}}$, $R_{K^{(\ast)}}$ Anomalies
}\\
\end{center}

\vspace{0.5cm}
\renewcommand{\thefootnote}{\fnsymbol{footnote}}
\begin{center}
{\large
{}~\textbf{Shaikh Saad}\footnote{ E-mail: \textcolor{MyLightBlue}{shaikh.saad@okstate.edu}}
 and 
{}~\textbf{Anil Thapa}\footnote{ E-mail: \textcolor{MyLightBlue}{thapaa@okstate.edu}} 
}
\vspace{0.5cm}

{\em Department of Physics, Oklahoma State University, Stillwater, OK 74078, USA }
\end{center}

\renewcommand{\thefootnote}{\arabic{footnote}}
\setcounter{footnote}{0}
\thispagestyle{empty}
\begin{abstract}
In this work, we present a solution to the persistent tensions in the decay observables $R_{D^{(\ast)}}$ and $R_{K^{(\ast)}}$ by introducing a $SU(2)_L$ doublet and a $SU(2)_L$ triplet scalar leptoquarks (LQs) that reside at the TeV energy scale. Neutrinos that remain massless in the Standard Model receive naturally small masses at one-loop level via the propagation of the same LQs inside the loop.  Such a common origin of apparently disjoint phenomenological observations is appealing, and we perform a comprehensive analysis of this set-up. We identify the minimal Yukawa textures required to accommodate these flavor anomalies and to successfully incorporate neutrino oscillation data while being consistent with all experimental constraints. This scenario has the potential to be tested at the experiments by the future improved measurements in the lepton flavor violating processes. Furthermore, proper explanations of these flavor anomalies predict TeV scale LQs that are directly accessible at the LHC.

\end{abstract}

\newpage
{\hypersetup{linkcolor=black}
\tableofcontents}
\setcounter{footnote}{0}
\section{Introduction}\label{SEC-01}
In the Standard Model (SM) of particle physics, all the charged fermions receive their masses as a result of the electroweak (EW) symmetry breaking, whereas neutrinos remain massless. Over the decades, neutrino oscillations have been firmly confirmed in the experiments \cite{Fukuda:1998mi, Ahmad:2002jz, Abe:2011sj, Adamson:2011qu, Abe:2011fz, An:2012eh, Ahn:2012nd} breaking the mold of the SM. However, the theoretical origin behind the neutrino mass generation is yet to be discovered. By extending the particle content of the SM, neutrino masses can be incorporated either at the tree-level via the seesaw mechanism \cite{Minkowski:1977sc, Yanagida:1979as, GellMann:1980vs, Mohapatra:1979ia, Schechter:1980gr, Schechter:1981cv, Foot:1988aq} or at the loop-level by quantum corrections \cite{Cheng:1977ir, Zee:1980ai, Cheng:1980qt, Zee:1985id, Babu:1988ki, Babu:1988ig, Ma:2006km}. We consider a scenario where neutrinos are Majorana fermions and their masses that violate lepton number by two units $\Delta L=2$ are generated radiatively at the one-loop order. In our set-up, due to loop suppression, neutrino masses are naturally small relative to the charged fermion masses. Scenarios, where the beyond SM  states running inside the loop have masses at the TeV scale are particularly interesting, as they can simultaneously solve other problems such as the recently observed flavor anomalies.   

Observation of neutrino oscillations provides direct evidence for flavor violation in the neutral lepton sector, which is indeed a phenomenon beyond the SM. Consequently, the origin of neutrino masses may be directly linked with flavor physics. Recently, several measurements associated with some of the flavor observables appear to indicate violations of lepton flavor universality (LFU). Four of such flavor observables, the ratios $R_{D^{(\ast)}}$ and  $R_{K^{(\ast)}}$, are of great importance. It is intrigued to find a common origin of the neutrino masses and $R_{D^{(\ast)}}$, $R_{K^{(\ast)}}$ anomalies, which is indeed the purpose of this work.   

Several independent measurements of semi-leptonic $B$ meson decays persistently show deviations from the SM predictions. Charged current processes lead to rare decays of the form $B\to D\tau \nu$ measured by Babar \cite{Lees:2012xj, Lees:2013uzd} and Belle \cite{Huschle:2015rga, Sato:2016svk, Hirose:2016wfn, Abdesselam:2019dgh}, as well as  $B\to D^{\ast}\tau \nu$ observed by Belle \cite{Huschle:2015rga, Sato:2016svk, Hirose:2016wfn, Abdesselam:2019dgh}, Babar \cite{Lees:2012xj, Lees:2013uzd}, and LHCb \cite{Aaij:2015yra, Aaij:2017uff} collaborations. The world averages of these measurements correspond to:
\begin{align}
&R_D=\frac{\Gamma(\overline{B}\to D\tau\nu)}{\Gamma(\overline{B}\to D\ell\nu)}=
\begin{cases}
0.299\pm 0.003\;\;\;\tt{SM}\;\text{\cite{Na:2015kha, Aoki:2016frl}},\\
0.334\pm 0.031\;\;\;\tt{exp}\;\text{\cite{Amhis:2016xyh, Abdesselam:2019dgh, Belle2019, Belle2019b}}.
\end{cases}
\\
&R_{D^{\ast}}=\frac{\Gamma(\overline{B}\to D^{\ast}\tau\nu)}{\Gamma(\overline{B}\to D^{\ast}\ell\nu)}=
\begin{cases}
0.258\pm 0.005\;\;\;\tt{SM}\;\text{\cite{Bigi:2017jbd, Jaiswal:2017rve, Bernlochner:2017jka}},\\
0.297\pm 0.015\;\;\;\tt{exp}\;\text{\cite{Amhis:2016xyh, Belle2019, Belle2019b}}.
\end{cases}
\end{align}
These experimental results point towards a tension of  $\gtrsim 3\sigma$ from SM predictions. Since these ratios are largely insensitive \cite{Bernlochner:2017jka} to hadronic uncertainties, the corresponding SM calculations are reliable.  

On the other hand, neutral current processes are responsible for $B$ meson decays of the form $B\to K^{(\ast)}\ell \ell$, which violates LFU in the following ratios:
\begin{align}
R_K=\frac{\Gamma(\overline{B}\to \overline{K}\mu^+\mu^-)}{\Gamma(\overline{B}\to \overline{K}e^+e^-)},\;\;\; R_{K^{\ast}}=\frac{\Gamma(\overline{B}\to \overline{K}^{\ast}\mu^+\mu^-)}{\Gamma(\overline{B}\to \overline{K}^{\ast}e^+e^-)}.   
\end{align}
The LHCb combined results of Run-1 data with  $2 fb^{-1}$ Run-2 data finds:
\begin{align}
&R_K=
\begin{cases}
1.0003\pm 0.0001\;\;\;\tt{SM}\; \text{\cite{Bobeth:2007dw}},\\
0.846^{+0.06+0.016}_{-0.054-0.014}\;\;\;\tt{exp}\;\;\text{\cite{Aaij:2019wad}}, \;\tt{for\; 1.1\; GeV^2<q^2<6.0\; GeV^2}.
\end{cases}
\end{align}
Here the first uncertainty is statistical, the second one is systematic, and $q^2$ is the di-lepton invariant mass squared. This result implies a tension of $\gtrsim 2.5 \sigma$ between the theory and experiment. For $R_{K^{\ast}}$, the Belle collaboration finds values at low and high $q^2$ bins given by:
\begin{align}
&R_{K^{\ast}}=
\begin{cases}
1.00\pm 0.01\;\;\;\tt{SM}\;\text{\cite{Bordone:2016gaq}},\\
0.90^{+0.27}_{-0.21}\pm 0.10\;\;\;\tt{exp}\;\text{\cite{Abdesselam:2019wac}}, \;\tt{for\; 0.1\; GeV^2<q^2<8.0\; GeV^2},\\
1.18^{+0.52}_{-0.32}\pm 0.10\;\;\;\tt{exp}\;\text{\cite{Abdesselam:2019wac}},  \;\tt{for\; 15\; GeV^2<q^2<19\; GeV^2}.
\end{cases}
\end{align}
Whereas these values are in agreement with the SM, the previous measurements by LHCb find a significant departure from theoretical calculations. The measured values are given as follows:
\begin{align}
&R_{K^{\ast}}=
\begin{cases}
0.660^{+0.110}_{-0.070}\pm 0.024\;\;\;\tt{exp}\;\text{\cite{Aaij:2017vbb}}, \;\tt{for\; 0.045\; GeV^2<q^2<1.1\; GeV^2},\\
0.685^{+0.113}_{-0.069}\pm 0.047\;\;\;\tt{exp}\;\text{\cite{Aaij:2017vbb}},  \;\tt{for\; 1.1\; GeV^2<q^2<6.0\; GeV^2}.
\end{cases}
\end{align}
These results show $\gtrsim 2.5 \sigma$ deviation from SM predictions in both $q^2$ bins. Theory predictions of both $R_K$ and $R_{K^{\ast}}$ are completely trustworthy since hadronic uncertainties cancel out in these ratios. 

These experimental results have drawn much attention to the theory community over the last several years.  Accurate measurements of these observables have the great potential to hunt for new physics (NP) since the corresponding ratios are reliably computed within the SM.  In short, experimental data of these flavor observable ratios beg for extensions of the SM for their proper explanations. To incorporate these significant deviations from the SM values, leptoquark (LQ) solutions are extensively exploited in the literature  (see for example Refs. \cite{Dorsner:2013tla, Sakaki:2013bfa, Duraisamy:2014sna,  Hiller:2014yaa, Buras:2014fpa,Gripaios:2014tna, Freytsis:2015qca, Pas:2015hca, Bauer:2015knc,Fajfer:2015ycq,  Deppisch:2016qqd, Li:2016vvp, Becirevic:2016yqi,Becirevic:2016oho, Sahoo:2016pet, Bhattacharya:2016mcc, Duraisamy:2016gsd, Barbieri:2016las,Crivellin:2017zlb, DAmico:2017mtc,Hiller:2017bzc, Becirevic:2017jtw, Cai:2017wry,Alok:2017sui, Sumensari:2017mud,Buttazzo:2017ixm,Crivellin:2017dsk, Guo:2017gxp,Aloni:2017ixa,Assad:2017iib, DiLuzio:2017vat,Calibbi:2017qbu,Chauhan:2017uil,Cline:2017aed,Sumensari:2017ovu, Biswas:2018jun,Muller:2018nwq,Blanke:2018sro, Schmaltz:2018nls,Azatov:2018knx, Sheng:2018vvm, Becirevic:2018afm, Hati:2018fzc, Azatov:2018kzb,Huang:2018nnq, Angelescu:2018tyl, DaRold:2018moy,Balaji:2018zna, Bansal:2018nwp,Mandal:2018kau,Iguro:2018vqb, Fornal:2018dqn, Kim:2018oih, deMedeirosVarzielas:2019lgb, Zhang:2019hth, Aydemir:2019ynb, deMedeirosVarzielas:2019okf, Cornella:2019hct,Datta:2019tuj, Popov:2019tyc, Bigaran:2019bqv, Hati:2019ufv, Coy:2019rfr, Balaji:2019kwe, Crivellin:2019dwb, 
Altmannshofer:2020axr, Cheung:2020sbq,Saad:2020ihm}).

In this work, we propose a solution to the $R_{D^{(\ast)}}$ and  $R_{K^{(\ast)}}$ anomalies by introducing a $SU(2)_L$ doublet and a $SU(2)_L$ triplet scalar LQs, respectively. Under the SM gauge group $SU(3)_C\times SU(2)_L\times U(1)_Y$, the quantum numbers of these scalars are $R_2 (3,2,7/6)$ and $S_3(\overline{3},3,1/3)$. As already aforementioned, it is appealing to connect these LFU violating anomalies with the flavor violation in the neutral lepton sector originating from the observation of non-zero masses of the neutrinos. In our proposed model, neutrino gets its mass by quantum corrections, where the same LQs ($R_2$ and $S_3$) run through the loop. However, one requires at least one more beyond the SM state to complete such loop diagrams, while breaking the lepton number by two units. For the sake of minimality, we discard the possibility of fermionic extensions and instead extend the scalar sector by a  $SU(2)_L$ singlet, which is in the $\chi_1(3,1,2/3)$ representation. 

We perform a detailed investigation of this model to identify the minimal Yukawa parameters that are required to satisfy the above-mentioned semi-leptonic $B$ meson decay anomalies and neutrino oscillation data. In our phenomenological analysis, we take into account all essential experimental constraints applicable to our model. We show that such a non-trivial task can be successfully achieved with only a limited number of parameters. The scenario presented in this work is particularly attractive since the same parameter set that generates neutrino masses and mixings, as well as accommodates semi-leptonic $B$ meson decay anomalies, also links many other beyond SM processes. Among them the charged lepton flavor violating processes that receive large contributions provide a way to probe this scenario in the experiments.  This model can also be tested at the LHC since existence of TeV scale LQs are required for consistent explanations of flavor anomalies.

Extension of the SM by scalars can be further motivated by their contributions toward the running of the SM Higgs self-coupling. It is well known that the SM vacuum is metastable  \cite{Holthausen:2011aa, EliasMiro:2011aa, Xing:2011aa, Alekhin:2012py}. This happens because the Higgs quartic coupling turns negative at large energy scales. Presence of new physics may provide sufficient  corrections to this running and make the universe stable.  The added scalars ($R_2, S_3, \chi_1$) within our scenario will provide non-zero corrections to the running of the SM Higgs self-coupling  via their Higgs portal  couplings, which at the one-loop level are always positive that helps to stabilize the potential \cite{Bandyopadhyay:2016oif}.   For detailed analysis including two-loop improved renormalization group equations running involving scalar LQ extension of the SM,  see for example Ref. \cite{Bandyopadhyay:2016oif}.

There are few more flavor observables that depart from their  theoretical  values. Associated to charged current processes, the ratio $R_{J/\psi}$ shows enhancement \cite{Aaij:2017tyk} in the tauoniuc mode to the muonic mode for $B\to J/\psi\ell\nu$, the longitudinal polarization of $D^{\ast}$ meson in the $B\to D^{\ast}\tau \nu$ decay known as  $f^{D^{\ast}}_L$ deviates from the SM value by $\sim 1.5\sigma$ \cite{Abdesselam:2019wbt}, and $\mathcal{P}^{\ast}_{\tau}$  the polarization asymmetry  in the longitudinal direction of the tau in the $D^{\ast}$ mode is also found to show some tension with theory \cite{Hirose:2016wfn}. These anomalies can also play prominent roles in finding NP, however, currently they have either large error bars or suffer from large hadronic uncertainties. This is why in this work we focus only on the  anomalies associated with $R_{D^{(\ast)}}$ and  $R_{K^{(\ast)}}$ ratios that provide clean signals of NP.

In Sec. \ref{SEC-02}, we introduce the proposed model. We discuss how to accommodate B physics anomalies and neutrino oscillation data in in Sec.~\ref{SEC-03} and Sec.~\ref{SEC-04}, respectively. Then we discuss at length all the experimental constraints that are relevant to our model in Sec.~\ref{SEC-05}. Furthermore, we present our results in Sec.~\ref{SEC-06}, and finally conclude in Sec. \ref{SEC-07}.

\section{The Model}\label{SEC-02}
In addition to the SM particle content, our proposed model consists of two scalar LQs $R_2(3,2,7/6)=\{R^{5/3}, R^{2/3}\}$ and  $S_3(\overline{3},3,1/3)=\{S^{4/3}, S^{1/3} S^{-2/3}\}$ to explain the $R_{D^{(\ast)}}$ and  $R_{K^{(\ast)}}$ anomalies, respectively. The corresponding Yukawa part of the Lagrangian is given by \cite{Buchmuller:1986zs}:
\begin{align}
    -\mathcal{L} \supset y_{ij}^L \overline{u}_{R i} R_2 i\sigma_2 L_{L_{j}} \,-\, y_{ij}^R \overline{Q}_{L i} R_2\ell_{Rj}\, + \, y_{ij} \overline{Q^c}_{L i} i\sigma_2 (\sigma^a S_3^a) L_{L j} + \text{h.c.},
    \label{LY}
\end{align}
where $Q_L$ and $L_L$ are the left-handed quark and lepton doublets under $SU(2)_L$, while $d_R$, $u_R$, and $\ell_R$ are  the right-handed down-type quark, up-type quark, and charged lepton, respectively that are singlets of $SU(2)_L$. Furthermore, $\sigma^a, a = 1,2,3$, are Pauli matrices, \{i,j\} are  flavor indices, and $S_3^a$ are the components of the LQ triplet. It is to be noted that $S_3$ couples to di-quarks, and we choose these couplings to be zero to ensure the stability of the proton. By expanding Eq. \eqref{LY} one can write: 
\begin{eqnarray}
    \mathcal{L}_{R_2}^{5/3} &=& - y_{i j}^L \overline{u}_{R i} \ell_{L j} R^{5/3} + (V y^R)_{i j} \, \overline{u}_{Li} \ell_{Rj} R^{5/3} + \text{h.c.}, \\
    \mathcal{L}_{R_2}^{2/3} &=& (y^L U)_{i j} \overline{u}_{Ri} \nu_{Lj} R^{2/3}  + y_{i j}^R \overline{d}_{Li} \ell_{Rj} R^{2/3} + \text{h.c.}, \\
     \mathcal{L}_{S_3} &=& (y U)_{i j} \overline{d^c}_{Li} S^{1/3} \nu_{Lj} + \sqrt{2} y_{i j} \overline{d^c}_{Li} S^{4/3} \ell_{Lj} + \sqrt{2} (V^* y U)_{i j} \overline{u^c}_{Li} S^{-2/3} \nu_{Lj} \nonumber\\ 
     &&- (V^* y)_{i j} \overline{u^c}_{Li} S^{1/3} \ell_{Lj} + \text{h.c.}. 
     \label{LYprime}
\end{eqnarray}
In writing the above Lagrangian terms, we have made a transition from flavor to mass eigenstates of the fermions with the following transformations:
\begin{align}
    \nu_L \to U \nu_L, \hspace{5mm} u_L \to V^\dagger u_L, \hspace{5mm} d_L \to d_L, \hspace{5mm} \ell_L \to \ell_L,
\end{align}
such that $V$ and $U$ are the Cabibbo-Kobayashi-Maskawa (CKM) and Pontecorvo-Maki-Nakagawa-Sakata (PMNS) matrices, respectively. We have chosen a basis where all the relative rotations are assigned to up-type quarks and neutrinos. From hereafter we will use the notation $\hat{y}^L_{ij}=(y^LU)_{ij}$. 

\begin{figure}[t!]
    \centering
    \includegraphics[scale=0.5]{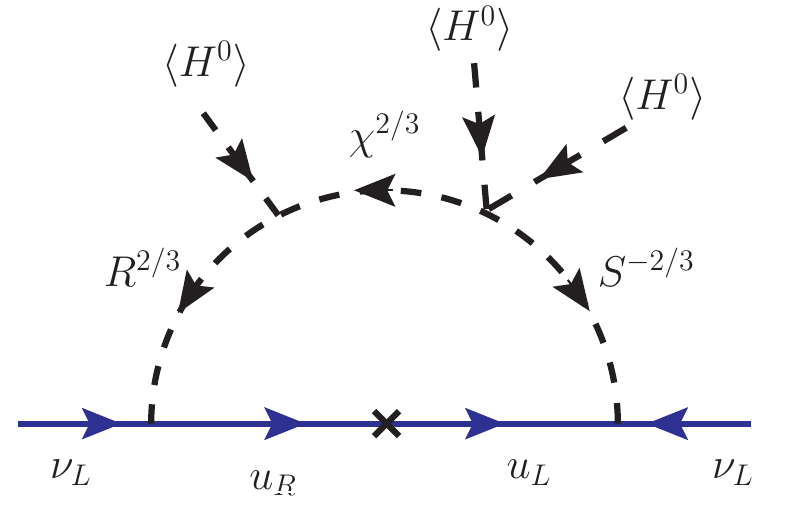}\hspace{0.2cm}
    \includegraphics[scale=0.5]{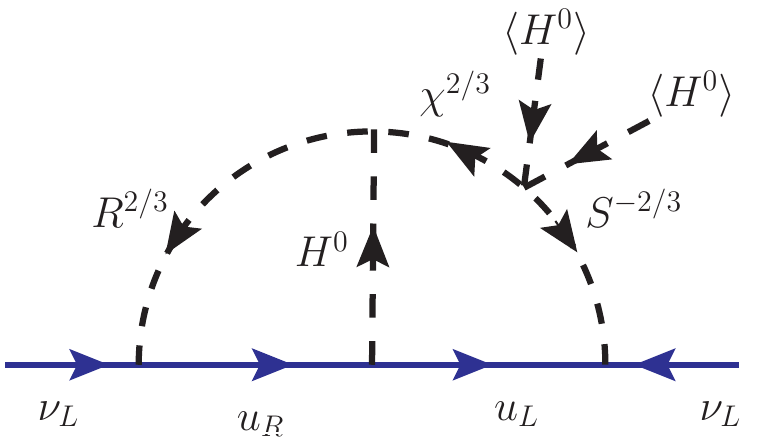}
    \caption{Representative Feynman diagrams of neutrino masses at the one-loop (left) and two-loop (right) levels.} 
    \label{fig:numass}
\end{figure}

The main objective in this work is to construct a minimal model that can simultaneously satisfy both the neutrino masses and the $B$ meson decay anomalies. However, with only $R_2$ and $S_3$ LQs, neutrino masses cannot be generated. For simplicity, instead of fermionic extensions, we prefer to enlarge the scalar sector. One such simple choice is to extend the particle content by a color triplet scalar  $\chi_1(3,1,2/3)=\chi^{2/3}$ \cite{Babu:2019mfe}, which is a singlet under $SU(2)_L$. Then in our model, neutrinos are Majorana type particles, and their masses are generated at the one-loop level, as shown in Fig. \ref{fig:numass}. In general, $\chi_1$ can have only di-quark coupling, which we assume to be sufficiently small. The completion of the one-loop diagram for neutrino mass requires the following two terms in the scalar potential:
\begin{align}
V\supset \mu \chi_1 H R^{\ast}_2 +\lambda H^{\ast}H^{\ast} \epsilon S_3 \chi_1 +h.c. \supset \mu\frac{v}{\sqrt{2}}\chi^{2/3}(R^{2/3})^{\ast}+\sqrt{2}\lambda  \frac{v^2}{2}\chi^{2/3}S^{-2/3}.    
\end{align}
Here $H$ is the SM Higgs doublet, $v=\langle H \rangle=246$ GeV, and $\epsilon$ is the rank-2 Levi-Civita tensor.

The one-loop neutrino mass matrix in this set-up is then given by:
\begin{align}
\mathcal{M}^\nu_{ij} = m_0\left[ (y^L)^{\ast}_{ki}m^u_k (V^{\ast}y)_{kj} + (i\leftrightarrow j) \right]
;\;\;\;m_0\approx \frac{1}{16 \pi^2}\frac{\mu\lambda v^3}{M^2_1M^2_2}, \label{numat}
\end{align}
here $m^u_{1,2,3}=\{m_u, m_c, m_t\}$, and $M_{1,2}$ are the two largest masses among $R^{2/3}, S^{-2/3}, \chi^{2/3}$ states. We denote the mass of the field $\chi$ by $m_{\chi}$, and the components of $R_2$ ($S_3$) LQ are assumed to be degenerate with a common mass  $m_{R}$ ($m_{S}$). Note that within this model, neutrino masses also receive contributions from a two-loop diagram, as shown in Fig. \ref{fig:numass}, which is expected to be much smaller than the one-loop contributions due to a further loop suppression; hence we neglect such corrections. Inclusions of such terms do not change the results presented in this work.

\begin{figure}[b!]
    \centering
    \includegraphics[scale=0.5]{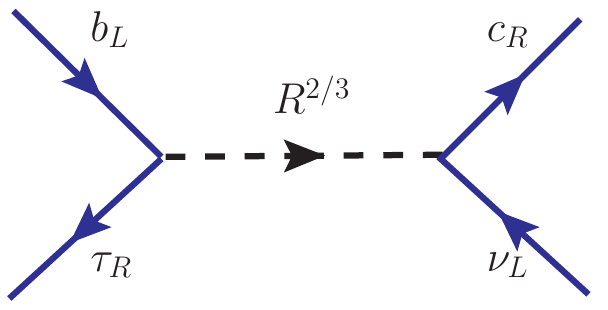}\hspace{2cm}
    \includegraphics[scale=0.5]{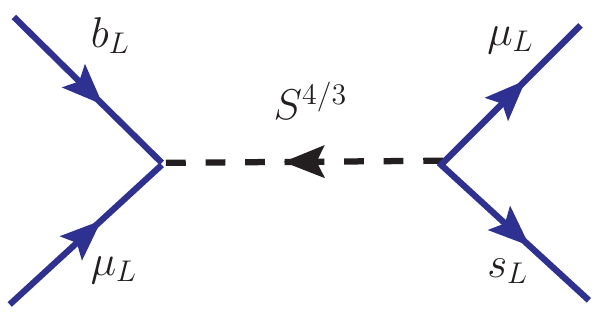}
    \caption{Representative Feynman diagrams that lead to $b\to c\tau \overline{\nu}$ (left) and $b\to s\mu^-\mu^+$ (right) transitions.}
    \label{fig:RDRK}
\end{figure}

\section{Explaining the B Meson Decay Anomalies}\label{SEC-03}
In this section, we discuss how to incorporate B meson decay anomalies observed in the  $R_{D^{(\ast)}}$ and  $R_{K^{(\ast)}}$ ratios within our framework.  
\subsection{\texorpdfstring{$R_D$-$R_{D^{\ast}}$  Anomalies}{RDD}}
Deviations that have been observed in several experiments in the  $R_D$ and $R_{D^{\ast}}$ ratios can be accommodated by the scalar LQ $R_2 (3,2,7/6)$. The charged current process $b\to c\tau \overline{\nu}$ contributing to $B\to D\tau\nu$ and $B\to D^{\ast}\tau\nu$ decays after integrating out the heavy LQ state $R_2$ can be explained in terms of the following effective Hamiltonian:
\begin{align}
\begin{aligned}
\mathcal{H}_{\mathrm{eff}}=\frac{4 G_{F}}{\sqrt{2}} V_{c b}\left[\left(\overline{\tau}_{L} \gamma^{\mu} \nu_{L\tau}\right)\left(\overline{c}_{L} \gamma_{\mu} b_{L}\right)\right.&+C^j_{S}(\mu)\left(\overline{\tau}_{R} \nu_{Lj}\right)\left(\overline{c}_{R} b_{L}\right) \\
&\left.+C^j_{T}(\mu)\left(\overline{\tau}_{R} \sigma^{\mu \nu} \nu_{Lj}\right)\left(\overline{c}_{R} \sigma_{\mu \nu} b_{L}\right)\right]+\mathrm{h.c.}, \label{ST}
\end{aligned}
\end{align}
where $C^j_{S,T}$ are the Wilson coefficients induced by the LQ state mediating the semileptonic decays via a tree-level contribution, as shown in Fig. \ref{fig:RDRK} (left diagram). The first term in the above Hamiltonian is the SM contribution, which strictly corresponds to $\nu_{\tau}$. On the other hand,  the BSM contributions need not conserve lepton flavor, and the neutrino index $j$ can in principle run over all three flavors \cite{Dorsner:2013tla, Sakaki:2013bfa}. Note that    experiments do not identify the flavor state of the neutrinos in the $R_D$-$R_{D^{\ast}}$ measurements.  After integrating out $R_2$ field, the expressions of the Wilson coefficients $C^j_{S,T}$ at the matching scale $\mu = m_R$ reads: 
\begin{align}
C^j_{S}\left(\mu=m_{R}\right)=4 C^j_{T}\left(\mu=m_{R}\right)=\frac{ \hat{y}^L_{c j} (y_{b \tau}^{R})^{\ast}}{4 \sqrt{2} m_{R}^{2} G_{F} V_{c b}}. \label{WC}
\end{align}

Here we comment on the scale dependence of the scalar and tensor Wilson coefficients. The corresponding hadronic operators given in Eq. \eqref{ST} have QCD anomalous dimensions, and their  associated matrix elements depend  on the renormalization scale. This renormalization scale dependence must be canceled in the observables by the opposite dependence of the Wilson coefficients. We are interested in the Wilson coefficients at the bottom-quark mass scale ($m_b=4.18$ GeV) at which the matrix elements of hadronic currents are calculated, which at the leading logarithm approximation have the following forms Ref.~\cite{Chetyrkin:1997dh, Gracey:2000am, Dorsner:2013tla, Hiller:2016kry} 
\begin{align}
&C_S(\mu=m_b)=\left[ \frac{\alpha_S(m_b)}{\alpha_S(m_t)}  \right]^{-\frac{\gamma_S}{2\beta^{(5)_0}}} \left[ \frac{\alpha_S(m_t)}{\alpha_S(m_R)}  \right]^{-\frac{\gamma_S}{2\beta^{(6)_0}}}  C_S(\mu=M_R),
\\
&C_T(\mu=m_b)=\left[ \frac{\alpha_S(m_b)}{\alpha_S(m_t)}  \right]^{-\frac{\gamma_T}{2\beta^{(5)_0}}} \left[ \frac{\alpha_S(m_t)}{\alpha_S(m_R)}  \right]^{-\frac{\gamma_T}{2\beta^{(6)_0}}}  C_T(\mu=M_R),
\end{align}
where the running coefficient is  $\beta^{(n_f)}_0=11-2 n_f/3$, with $n_f=$ number of active quark flavors,  and the coefficients of the anomalous dimensions are given by $\gamma_S=-8$ and $\gamma_T=8/3$.  Whereas the above analytical formulas can be straightforwardly used to find the running of the Wilson coefficients, however, higher order QCD corrections can be significant, and need to be solved numerically. The corresponding QCD  anomalous dimensions up-to three loop are well known and are listed in Ref. \cite{Gonzalez-Alonso:2017iyc}, along with the relevant EW anomalous dimensions. In our numerical analysis we utilize {\tt Flavio} package \cite{Straub:2018kue} that has NNLO QCD and NLO EW corrections coded in it. We find the correlations between the Wilson coefficients at the two different scales to be $C_S(m_b)= 1.628 \;C_S(M_R)$  and $C_T(m_b)= \;0.857 C_T(M_R)$ (here we have taken $m_R=1$ TeV). These lead to $C^j_S(m_b)= 7.6\; C^j_T(m_b)$, which breaks the relation at the LQ mass scale given in Eq. \eqref{WC}.  Here for simplicity, we have omitted the mixing terms between the scalar and the tensor Wilson coefficients. However, this approximation  is valid to a very good degree since the off diagonal mixing terms are sufficiently small compared to the diagonal entries, see Ref. \cite{Gonzalez-Alonso:2017iyc} for details. Throughout this work, we will quote the values of these ($C_S, C_T$) Wilson coefficients at the scale of the heavy NP particle.

\begin{figure}[t!]
    \centering
    \includegraphics[scale=0.27]{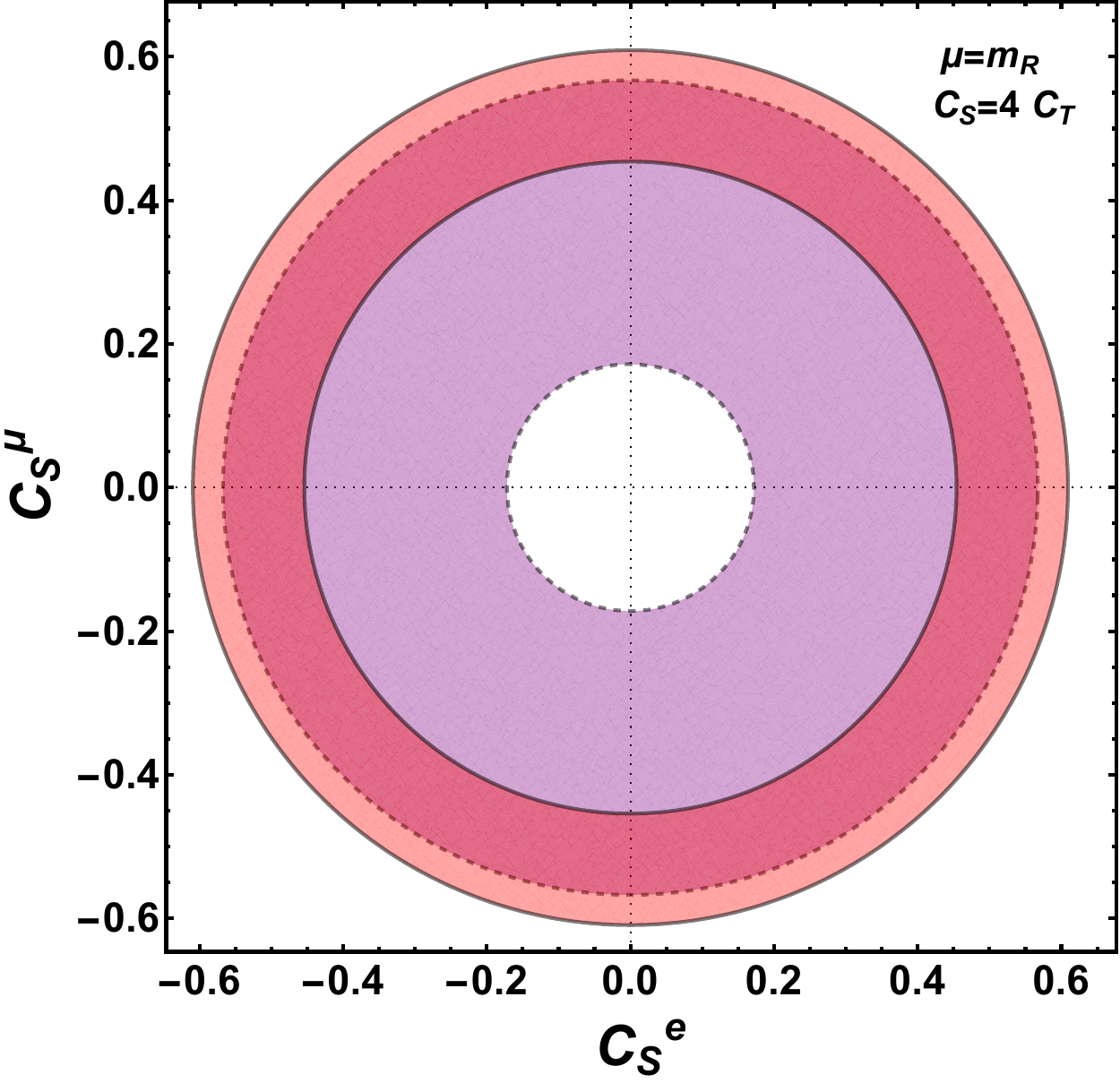}\hspace{2cm}
    \includegraphics[scale=0.27]{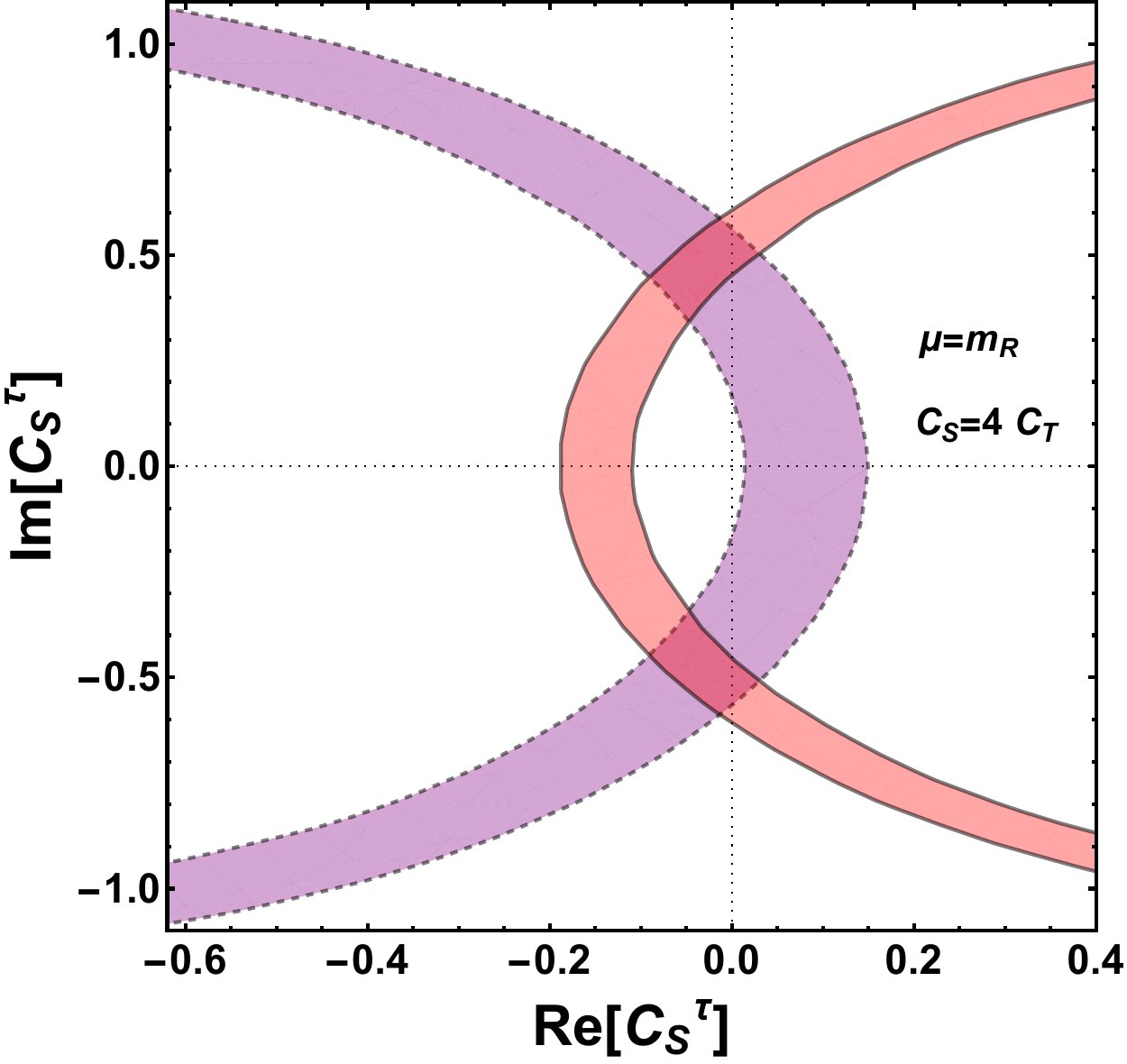}
    \caption{Fits to the $R_D$-$R_{D^{\ast}}$ in a two parameter scenario  with (i) $C^{\tau}_L=0$,  $C^{e,\mu}_L\in \mathcal{R}$, and (ii) $C^{e,\mu}_L=0$, $C^{\tau}_L\in \mathcal{C}$. The purple (pink) band represents $R_D$ ($R_{D^{\ast}}$) within experimental $1\sigma$ values. We are interested in the values of the corresponding Wilson coefficients  in the  overlapping shaded regions  between these two bands. }
    \label{fig:RDfit}
\end{figure}

From the discussion above, it is clear that a successful explanation of the $R_D$-$R_{D^{\ast}}$ anomalies requires the coupling of the $b$-quark to $\tau$-lepton. Moreover, $c$-quark must couple to neutrinos $\nu_j$, implicating the following needed Yukawa textures \cite{Dorsner:2013tla}:
\begin{align}
y^R=\left(\begin{array}{lll}
0 & 0 & 0 \\
0 & 0 & 0 \\
0 & 0 & y_{33}^R
\end{array}\right),\;\; 
\hat{y}^L = \left(\begin{array}{lll}
0 & 0 & 0 \\
\hat{y}_{21}^L & \hat{y}_{22}^L & \hat{y}_{23}^L \\
0 & 0 & 0
\end{array}\right). \label{YmatRD}
\end{align}
We remind the readers that the $\hat{y}^L$ matrix is related to the original $y^L$ matrix via PMNS rotation: $\hat{y}^L=y^LU$.
With this form of the Yukawa coupling matrices, a solution to the $R_D$-$R_{D^{\ast}}$ may be achieved in a few different ways: (i) by considering $j=\tau$ in Eq. \eqref{WC}, where a constructive interference with the SM semileptonic amplitude occurs, (ii) by considering $j\neq \tau$, where no such interference is realized, and (iii) by considering a mixed scenario where all neutrino species contribute.  The required values of the Wilson coefficients to simultaneously fit $R_D$ and $R_{D^{\ast}}$ anomalies in the cases of two-parameter scenarios are depicted in Fig. \ref{fig:RDfit}.  The left plot corresponds to the case with real $C^{e,\mu}_S$ ($C^{\tau}_S=0$), and the right plot represents the scenario with  complex $C^{\tau}_S$ ($C^{e,\mu}_S=0$). We have used {\tt Flavio} package to get these plots.  We find that scenarios, where $j=\tau$  contribution is non-negligible toward $R_{D^{(\ast)}}$, it always needs to be very close to purely imaginary to get a good fit. For scenarios of this type, we choose $y^R_{33}$ to be purely imaginary.  On the other hand, all  parameters can be taken to be real for cases when $j=\tau$  contribution is negligible (as can be seen from the left plot in Fig. \ref{fig:RDfit}). 

As well known, $S_3$ LQ can also contribute to $b\to c\tau \overline{\nu}$ process with a wrong sign that is at odds with the experimental observation. These contributions are proportional to the product of Yukawa couplings $y_{3j}(Vy^{\ast})_{23}$, which within our set-up is too small to provide any significant contribution.

\subsection{\texorpdfstring{$R_K$-$R_{K^{\ast}}$ Anomalies}{RKK}}
The $B$ meson decay anomalies associated with the neutral current processes correspond to an excess of $B\to K^{(\ast)} \mu^+ \mu^-$ over the  $B\to K^{(\ast)} e^+ e^-$ channels. The  effective Hamiltonian that describes these processes can be written as follows:
\begin{align}
\mathcal{H}_{eff}=-\frac{4 G_F}{\sqrt{2}} V_{tj}V^{\ast}_{ti}\left(\sum_{X=9,10}C_X^{ij,\ell\ell^{\prime}} \mathcal{O}_X^{ij,\ell\ell^{\prime}}\right)  +h.c. \label{Heff} 
\end{align}
Here the effective operators correspond to:
\begin{align}
\mathcal{O}_9^{ij,\ell\ell^{\prime}}=\frac{\alpha}{4\pi}\left(\overline{d}_i\gamma^{\mu}P_Ld_j\right)\left(\overline{\ell}\gamma_{\mu}\ell^{\prime}\right),\;
\mathcal{O}_{10}^{ij,\ell\ell^{\prime}}=\frac{\alpha}{4\pi}\left(\overline{d}_i\gamma^{\mu}P_Ld_j\right)\left(\overline{\ell}\gamma_{\mu}\gamma_5\ell^{\prime}\right).
\end{align}
Within our framework, these types of effective operators are generated at the tree-level via  $S_3$ LQ, as presented in Fig. \ref{fig:RDRK} (right diagram). After integrating out the heavy degrees of freedom, one gets the following Wilson coefficients:
\begin{align}
C^{\ell\ell^{\prime}}_{9}=-C^{\ell\ell^{\prime}}_{10}=\frac{v^{2}}{V_{t b} V_{t s}^{\ast}}\frac{\pi}{\alpha_{em}} \frac{y_{b\ell^{\prime}} \left(y_{s \ell} \right)^{*}}{m_{S}^{2}}.
\end{align}
Wilson coefficients $C_{9,10}$ get additional contributions from the one-loop mediated processes lead by $R_2$ LQ, which can be safely ignored in the present scenario. Interestingly, $S_3$ is the only LQ that can explain both $R_K<R^{SM}_K$ and $R_{K^{\ast}}<R^{SM}_{K^{\ast}}$ at the tree-level.

In this work, we assume the NP dominantly couples to the muon sector (hence $\ell=\ell^{\prime}=\mu$), and assume the following texture of the corresponding Yukawa matrix: 
\begin{align}
y = \left(\begin{array}{lll}
0 & 0 & 0 \\
0 & y_{22} & 0 \\
0 & y_{32} & 0
\end{array}\right), \label{YmatRK}
\end{align}
to alleviate discrepancies in the $R_{K^{(\ast)}}$ ratios.  The scenario with $C^{\mu\mu}_{9,10}$ is shown to provide an excellent fit to the  $R_K$-$R_{K^{\ast}}$ anomalies, and the required values of the Wilson coefficients corresponding to best fit is given by \cite{Aebischer:2019mlg}:
\begin{align}
C^{\mu\mu}_{9}=-C^{\mu\mu}_{10}=-0.53. \label{c9}  
\end{align}
The allowed values within the one-sigma and two-sigma correspond to $\left[-0.61,-0.45\right]$ and $\left[-0.69,-0.37\right]$, respectively. It should be clarified that in addition to $R_{K^*}$ ratios, these are several more observables linked with the neutral current processes that show some moderate level of discrepancies compared to SM predictions; two important quantities  are the angular observable $P^{\prime}_5$ \cite{Aaij:2015oid, Khachatryan:2015isa, Aaboud:2018krd}, and combined fit to the anomalous $b\to s$ data in operators contributing to $b\to s\mu\mu$ \cite{DAmico:2017mtc, Geng:2017svp, Capdevila:2017bsm, Altmannshofer:2017yso, Ciuchini:2017mik, Hiller:2017bzc, Aebischer:2019mlg, Ciuchini:2019usw, Kowalska:2019ley, Alok:2019ufo}. Explanations of all these deviations in the neutral current processes are explained by the fit obtained in Eq. \eqref{c9}, for more details see Ref. \cite{Aebischer:2019mlg}.

\section{Incorporating Neutrino Oscillation Data}\label{SEC-04} 
In our proposed model, neutrinos are Majorana fermions, and as aforementioned, neutrino mass is generated radiatively at the one-loop level.
The computed neutrino mass matrix is given in Eq.~\eqref{numat} which can be decomposed into a from: $\mathcal{M}^{\nu}= U\; diag\{m_1,m_2,m_3\}\; U^T$. Whereas from experimental results, we already know the definite sign of $\Delta m^2_{21}>0$; however, the sign of $\Delta m^2_{31}$ is yet to be measured. The case with $\Delta m^2_{31}>0$ corresponds to the normal ordering (NO). In contrast, the case with $\Delta m^2_{31}<0$  is known as the inverted ordering (IO). Since inverted mass ordering is less preferred by fits to data, this work only considers normal mass ordering. The PMNS matrix $U$ consists of three physical mixing angles that are measured with excellent precision in the experiments. Furthermore, two of the mass squared differences have been measured with great accuracy, whereas the absolute scale of the neutrinos is yet to be determined. These experimental values are summarized in Table \ref{nupara}, which we use for numerical analysis. Since the Dirac phase in the neutrino sector is inadequately measured in the experiments, and the Majorana phases have not been measured, we do not include these quantities in our fits.

\begin{table}[b!]
\centering
\begin{tabular}{|c|c|c|}
\hline \hline
\textbf{Oscillation} & \textbf{1 $\sigma$ allowed range} & \textbf{3 $\sigma$ allowed range} \\
\textbf{parameters} &  \textbf{from NuFit4.1} & \textbf{from NuFit4.1} \\ \hline \hline
$\sin^2\theta_{12}$ & $0.31_{-0.012}^{+0.013}$ & 0.275 - 0.350 \\ \hline
$\sin^2\theta_{13}$  & $0.02241_{-0.00065}^{+0.00066}$ & 0.02046 - 0.02440 \\ \hline
$\sin^2\theta_{23}$  & $0.558_{-0.033}^{+0.020}$ & 0.427 - 0.609 \\ \hline
$\Delta m_{21}^2$ $( 10^{-5} eV^2)$ & $7.39_{-0.20}^{+0.21}$ & 6.79 - 8.01  \\ \hline
$\Delta m_{31}^2$ $( 10^{-3} eV^2)$  & $2.523_{-0.030}^{+0.032}$ & 2.432 - 2.618 \\ \hline
\end{tabular}
\caption{Neutrino oscillation parameters for the NO scenarios, these values are taken from recent NuFit collaboration \cite{Esteban:2018azc}. }\label{nupara}
\end{table}

By combining the Yukawa textures Eqs. \eqref{YmatRD} and \eqref{YmatRK}, along with the neutrino mass matrix given in Eq.~\eqref{numat}, it is trivial to verify that neutrino oscillation data cannot be explained. Hence to properly incorporate neutrino masses and mixings, we must enlarge the Yukawa parameters. Since in $y^L$ matrix, which enters the neutrino mass matrix formula and contains non-zero entries (entry) in the second row that are (is) required for the explanation of $R_{D^{(\ast)}}$, it is natural to populate along the second row of $y$ matrix as well, for the sake of minimality. Following this argument, our detailed numerical study performed in Sec.  \ref{SEC-06} reveals two minimal textures for the Yukawa coupling matrices to accommodate anomalies in the $R_{D^{\ast}}$, $R_{K^{\ast}}$ ratios in conjunction with neutrino observables, which are:      
\begin{align}
&\texttt{TX-I}:\;\;\;
y^R=\left(\begin{array}{lll}
0 & 0 & 0 \\
0 & 0 & 0 \\
0 & 0 & \blue{\ast}
\end{array}\right),\;\;
y^L = \left(\begin{array}{lll}
0 & 0 & 0 \\
\blue{\ast} & \blue{\ast} & \blue{\ast} \\
0 & 0 & \ast
\end{array}\right),\;\;
y = \left(\begin{array}{lll}
0 & 0 & 0 \\
\ast & \red{\ast} & \ast \\
0 & \red{\ast} & 0
\end{array}\right). \label{TX-I}
\\
&\texttt{TX-II}:\;\;\;
y^R=\left(\begin{array}{lll}
0 & 0 & 0 \\
0 & 0 & 0 \\
0 & 0 & \blue{\ast}
\end{array}\right),\;\;
y^L = \left(\begin{array}{lll}
0 & 0 & 0 \\
\blue{\ast} & \blue{\ast} & \blue{\ast} \\
0 & 0 & \ast
\end{array}\right),\;\;
y = \left(\begin{array}{lll}
0 & 0 & 0 \\
0 & \red{\ast} & \ast \\
\ast & \red{\ast} & 0
\end{array}\right). \label{TX-II}
\end{align}
Here the entries in blue enter the explanation of $R_D$, $R_{D^{\ast}}$ ratios, whereas entries in red play a role in explaining $R_K$, $R_{K^{\ast}}$ ratios. Hence, we always keep these terms to be non-zero. Additional entries shown in black allow fitting the neutrino observables as well. Moreover, in the above Yukawa coupling matrices, no entries for the first generation quarks are introduced, since they are highly constrained by experimental data. As will be shown in Sec. \ref{SEC-06}, with only a limited number of parameters as given in Eqs. \eqref{TX-I} and \eqref{TX-II} excellent fits to all the observables can be achieved. Neutrino observables, as well as  $R_{D^{\ast}}$, $R_{K^{\ast}}$ ratios can all be fitted within their $1\sigma$ experimental values. A choice of such Yukawa couplings texture are subjected to various experimental constraints that we discuss at length in the next section. It is worthwhile to mention that further reduction of the number of parameters in Eqs. \eqref{TX-I} and \eqref{TX-II} do not return the aforementioned observables within $1\sigma$ measured values. Hence we do not list or consider such textures here.

\section{Experimental Constraints}\label{SEC-05}
In this section, we summarize all the relevant experimental constraints associated with the BSM states $R_2$ and $S_3$.   For the phenomenological analysis, we limit ourselves to the case of no mixing among the different LQs. Small mixings among $R^{2/3}, S^{2/3}$, and $\chi^{2/3}$ are indeed required to get the correct scale for the neutrino masses, as will be clarified later in the text (see Sec. \ref{SEC-06}).

\subsection{Radiative Leptonic Decays: \texorpdfstring{$\ell\to \ell^{\prime}\gamma$}{ellell}}
LQ mediated one-loop contributions to the dangerous lepton flavor violating (LVF) processes $\ell\to \ell^{\prime}\gamma$ can severely constrain the couplings of the LQ with the charged leptons. In our model, some of the parameters are highly constrained by these radiative decay processes. Following the general method described in Ref.~\cite{Lavoura:2003xp} (see also Ref.~\cite{Dorsner:2016wpm}) in computing such processes, we derive the following contributions to LVF decays within our scenario: 
\begin{align}
&Br\left(\ell\to \ell^{\prime}\gamma\right)=\frac{\tau_{\ell}\;9\alpha\; m^5_{\ell}}{4(16\pi^2)^2}
\nonumber \\ &\times
\left|\sum_{q} \left( 
\frac{y^L_{q\ell}(y^L_{q\ell^{\prime}})^{\ast}}{4m^2_R}
+\frac{(Vy^R)_{q\ell}(y^L_{q\ell^{\prime}})^{\ast}}{m^2_R} \frac{m_q}{6m_{\ell}}\left(1+4\log x_q\right) + \frac{y^{\ast}_{q\ell^{\prime}}y_{q\ell}}{3m^2_{S}}  - \frac{(V^{\ast}y)_{q\ell^{\prime}}(V^{\ast}y)_{q\ell}}{12m^2_{S}}
\right)
\right|^2.
\end{align}
Here $\tau_{\ell}$ is the lifetime of the initial state lepton, and $x_q=(m_q/m_{LQ})^2$. In the above formula, the first (last) two terms are the contributions from the $R_2$ ($S_3$) LQ and the expression is in agreement with Ref. \cite{Becirevic:2017jtw} (\cite{Hati:2018fzc}).   The current experimental bounds on these processes are given by \cite{ TheMEG:2016wtm, Aubert:2009ag}:
\begin{align}
&Br\left(\mu\to e\gamma\right)<4.2\times 10^{-13},\\ 
&Br\left(\tau\to e\gamma\right)<3.3\times 10^{-8}, \\
&Br\left(\tau\to \mu\gamma\right)<4.4\times 10^{-8},
\end{align}
which can be easily translated into the upper bounds on the Yukawa couplings and the lower bound on the mass of the corresponding LQ. 
We find that, as long as these flavor violating processes are under control, processes like  $\ell\to \ell^{\prime}\ell^{\prime}\ell^{\prime}$ are well below their current experimental bounds. Even though in our numerical analysis, we take into account all such processes, here for brevity, we do not write down the associated formulas.      

\subsection{\texorpdfstring{$\mu - e$}{mue} Conversion} 
Neutrinoless $\mu - e$ conversion in nuclei can provide strong bounds on some of the Yukawa couplings in our model. The LQ $S_3$  mediates this dangerous process at the tree-level, and the associated Yukawa couplings are directly linked to neutrino observables and anomalies in the $R_{K^{(\ast)}}$ ratios.   This conversion rate can be written as follows:     
\begin{align}
\text{CR}(\mu-e) 
= \frac{\Gamma^{\mu-e} }{\Gamma_\text{capture}(Z)}.
\end{align}
Here $\Gamma_\text{capture}(Z)$ represents the capture rate for a nucleus with atomic number $Z$, and the expression for the  width 
$\Gamma^{\mu-e}$ in our scenario is given by \cite{Kitano:2002mt,Dorsner:2016wpm}:
\begin{align}\label{mue}
\Gamma^{\mu-e} \,= \,2 \,G_F^2\, 
\left| (2 V^{(p)} + g_{LV}^{(u)} V^{(n)}) g_{LV}^{(u)}\right|^2;\;\;\;
g_{LV}^{(u)}=\frac{-2v^2}{m_{S_3}^2} \,  
(V^{\ast} y)_{u\ell'}\, (V^{\ast}y)^*_{u\ell}.
\end{align}
In Eq. \eqref{mue} the nuclear form factors (in units of $m_{\mu}^{5/2}$) are given by \cite{Kitano:2002mt}  $V^{(p)}=0.0974$,  $V^{(n)}=0.146$ for gold and the total capture rate for gold is $\Gamma_{\text{capture}}=13.07\times 10^6$ $s^{-1}$. The current  sensitivity imposes strong bounds on the $\mu - e$ transition rate: $\text{CR}(\mu-e)<7\times 10^{-13}$   \cite{Bertl:2006up}.  On the other hand,  the future projected sensitivity  that will provide an improvement on the current limit by almost four orders
of magnitude $CR(\mu\to e)< 10^{-16}$  \cite{Kurup:2011zza, Cui:2009zz, Chang:2000ac, Adamov:2018vin, Bartoszek:2014mya, Pezzullo:2018fzp, Bonventre:2019grv} is capable of ruling out  most part of the parameter space of this theory.

\subsection{\texorpdfstring{$Z$-Boson Decays: $Z\to \ell \ell^{\prime}$}{Zellell}}
Leptoquarks' contributions to $Z$ decays to pair of leptons via one-loop processes provide substantial restrictions on the size of the Yukawa couplings. In the following, we derive such constraints on the relevant Yukawa couplings of the LQs. The effective Lagrangian of the following form can describe these processes \cite{Arnan:2019olv}:
\begin{align}
\delta \mathcal{L}^Z_{eff}=\frac{g}{\cos \theta_W} \sum_{f,i,j} \overline{f}\gamma^{\mu} \left( g^{ij}_{f_L}P_L + g^{ij}_{f_R}P_R \right) f_j Z_{\mu},   
\end{align}
with $g^{ij}_{f_{L(R)}}=\delta_{ij}g^{SM}_{f_{L(R)}} + \delta g^{ij}_{f_{L(R)}}$, $g^{SM}_{f_L}=I^f_3-Q^f \sin^2 \theta_W$, $g^{SM}_{f_R}=-Q^f \sin^2 \theta_W$,  and $\theta_W$ is the Weinberg angle. These dimensionless coupling parameters $g^{ij}$ are responsible for Z-decay processes $Z\to \ell \ell^{\prime}$.  LEP \cite{Tanabashi:2018oca} measurements provide stringent constraints on the contributions to these rates for both flavor conserving and flavor violating $Z$ decays. By following the computations done in Ref.~\cite{Arnan:2019olv}, we derive the contributions by the  LQs to $Z$ decays, which are given by:
\begin{align}
Re\left[ \delta g^\ell_{L,R} \right]^{ij} &=  \dfrac{3w^u_{tj}(w^u_{ti})^{\ast}}{16 \pi^2} \Bigg{[} (g_{u_{L,R}}-g_{u_{R,L}})\dfrac{x_t (x_t-1- \log x_t)}{(x_t-1)^2} \Bigg{]}\nonumber \\
&+ \dfrac{x_Z}{16\pi^2} \sum_{q=u,c}w^u_{qj}(w^u_{qi})^{\ast}  \Bigg{[} -g_{u_{R,L}}\left( \log x_Z-\frac{1}{6} \right)+\frac{g_{\ell_{L,R}}}{6} \Bigg{]}\nonumber \\
&+\dfrac{x_Z}{16\pi^2} \sum_{q=d,s,b}w^d_{qj}(w^d_{qi})^{\ast}  \Bigg{[} -g_{d_{R,L}}\left( \log x_Z-\frac{1}{6} \right)+\frac{g_{\ell_{L,R}}}{6} \Bigg{]}, \label{gLR}
\end{align}
where $x_Z=m^2_Z/m^2_{LQ}$, $x_t=m^2_t/m^2_{LQ}$ and $g_{f_L}= I^f_3-Q^f\sin^2\theta_W$, $g_{f_R}=-Q^f\sin^2\theta_W$. Furthermore, for LQ $R_2$ we have defined: $w^u_{ij}=-y^L_{ij}$, $w^d_{ij}=0$ when calculating $\delta g_L$. For the same LQ, while calculating $\delta g_R$  we have: $w^u_{ij}=(Vy^R)_{ij}$, $w^d_{ij}=y^R_{ij}$.  On the other hand, for $S_3$ LQ, $w^u_{ij}=-(V^{\ast}y)_{ij}$, $w^d_{ij}=-\sqrt{2}y_{ij}$ replacements must be made, and this LQ does not give any contribution to $\delta g_R$. Also for $S_3$ LQ, in the second and third lines in Eq. \eqref{gLR}, a further replacement of $g_{{u,d}_{R,L}}\to g_{{u,d}_{L,R}}$ must be made.  Some of the Yukawa couplings are strongly constrained by the precise measurements of these  effective couplings by the LEP collaboration \cite{ALEPH:2005ab}: 
\begin{align}
&Re[\delta g^e_L]\leq 3.0\times 10^{-4},\;\;  Re[\delta g^\mu_L]\leq 1.1\times 10^{-3}, \;\;Re[\delta g^\tau_L]\leq 5.8\times 10^{-4},\\
&Re[\delta g^e_R]\leq 2.9\times 10^{-4},\;\;  Re[\delta g^\mu_R]\leq 1.3\times 10^{-3},\;\; Re[\delta g^\tau_R]\leq 6.2\times 10^{-4}.
\end{align}
These experimental results can be translated into upper bounds on the Yukawa couplings of our theory for a fixed value of LQ masses.  Moreover, LHC data \cite{Aad:2014bca} imposes an upper limit on the branching fraction, $B(Z\to e \mu)<7.5\times 10^{-7}$ at the $95\%$ confidence level. Within our scenario, contributions from $S_3$ can be safely neglected. Moreover, one of the essential bounds on the Yukawa couplings we get from the above LEP data is (with $m_{R}=1$ TeV):
\begin{align}
y^R_{33}\leq 0.856 (1\sigma) \;[1.22 (2\sigma)].
\end{align}
For our numerical study, we demand its value to lie within the $2\sigma$ range, which is crucial to achieving a good fit in our model. 

Additionally, the effective number of neutrinos is another  accurately measured quantity, and the observed value corresponds to \cite{ALEPH:2005ab}:
\begin{align}
N_\nu^{\mathrm{exp}} = 2.9840\pm 0.0082,
\end{align}
which will constraint the LQ couplings to the neutrinos by the following equation:
\begin{align}
N_\nu = \sum_{i,j} \left|\delta_{ij}+\dfrac{\delta g_{\nu_L}^{ij}}{g_{\nu_L}^{\mathrm{SM}}}\right|^2.
\end{align}
While performing a fit, we also incorporate these bounds on the Yukawa couplings.

\subsection{\texorpdfstring{$B_c$ Meson Lifetime: $B_c\to \tau \nu$}{MesonL}} 
The same Yukawa couplings relevant for $R_{D^{(\ast)}}$ contribute to the decay $B_c\to \tau \nu$. 
The lifetime of $B_c$ meson is a very precisely measured quantity \cite{Tanabashi:2018oca}: $\tau[B_c]=(0.507\pm 0.009)$ ps. This has a significant impact on physics BSM that leads to $B$ meson decay and constraints the yet unmeasured branching ratio $Br(B_c\to \tau \nu)$. In our scenario, this branching ratio has the form \cite{Watanabe:2017mip}:
\begin{align}
Br(B_c\to \tau \nu)=0.023 \left|1-4.068\; C_S(\mu=m_{B_c})\right|^2.
\end{align}
In deriving this formula, we have used $f_{B_c}=0.43$ GeV and $m_{B_c}=6.2749$ GeV. 

Since the branching ratio $Br(B_c\to \tau \nu)$ has not been measured yet in the experiments, $B_c$ lifetime needs to be compared with the theoretical calculations \cite{Gershtein:1994jw, Bigi:1995fs,Beneke:1996xe, Chang:2000ac,  Kiselev:2000pp}.   By doing so, the works in Ref. \cite{Akeroyd:2017mhr} and Ref. \cite{Alonso:2016oyd} advocated that the NP contributions to this decay must be $Br(B_c\to \tau \nu)\leq  10\%$ and $Br(B_c\to \tau \nu)\leq 30\%$, respectively. However, as claimed in Ref. \cite{Blanke:2018yud} (see also Ref.~\cite{Bardhan:2019ljo}), these calculations suffer from theoretical uncertainties, and suggested a conservative limit of $Br(B_c\to \tau \nu)\leq 60\%$. As will be shown, a somewhat larger branching ratio is required to be compatible with $R_{D^{(\ast)}}$ measurements in our set-up.

\subsection{Rare Meson Decays: \texorpdfstring{$B\to K^{(\ast)}\nu_{\ell}\nu_{\ell^{\prime}}$}{RareM}}
NP explanations of B-anomalies impose strong constraints on the couplings due to rare meson decays associated with processes like $s\to d\nu \nu$ and $b\to s\nu\nu$. Note that according to the specific choice of the Yukawa coupling given in Eqs. \eqref{TX-I} and \eqref{TX-II}, our set-up does not lead to the former mode; however the latter mode is open. This will correspond to $B\to K^{(\ast)}\nu_{\ell}\nu_{\ell^{\prime}}$ that can be computed from the effective Hamiltonian given in Eq.~\eqref{Heff}. Which in our case is mediated by $S_3$ LQ, and upon integrating out this heavy state, this process can be explained by \cite{Bobeth:2017ecx,Bordone:2017lsy}:
\begin{align}
\mathcal{H}_\text{eff}^\text{LQ}(d_j \to d_i \nu_\ell\bar
\nu_{\ell^\prime}) 
=  - \frac{(y U)_{j\ell^\prime} (y U)_{i\ell}^\ast}{2m_{S_3}^2} 
\left(\bar d_i \gamma_\mu  P_L  d_j\right) 
\left(\bar \nu_\ell \gamma^\mu  P_L \nu _{\ell^\prime}\right).
\end{align}
Then the corresponds Wilson coefficient is given by: 
\begin{align}
C_{bs}^{\ell\ell^{\prime},NP} = 
\frac{v^2}{V_{ts}^\ast V_{tb}} \frac{\pi}{\alpha_{em}}
\frac{(yU)_{s\ell}^\ast(yU)_{b\ell^\prime}}{2 m_{S_3}^2}.
\end{align}
Using these results one can write:
\begin{align}
R^{\nu\nu}_{K^{(\ast)}} = \frac{\Gamma_\text{SM + NP}(B\to
K^{(\ast)}\nu\nu)}{\Gamma_\text{SM}(B\to K^{(\ast)}\nu\nu)}=
\frac{1}{3 \,|C^{SM}_{bs}|^2}\sum_{\ell,\ell^\prime} 
\left|C_{bs}^{\ell\ell,\text{SM}}\,\delta_{\ell\ell^\prime}\,+\, 
C_{bs}^{\ell\ell^\prime,\text{NP}}\right|^2\,,
\end{align}
here $C_{\text{SM}}^{\ell\ell} = -6.38(6)$ (for each neutrino
flavour) \cite{Brod:2010hi}. From the above expressions, one can find limits on the Yukawa couplings and the mass of the associated LQ by comparing with the SM predictions \cite{Buras:2014fpa}: 
BR$(B^+\to K^+ \nu\bar\nu) = (4.0\pm 0.5) 
\times 10^{-6}$ and
BR$(B^0\to K^{\ast 0} \nu\bar \nu) = (9.2\pm 1.0) 
\times 10^{-6}$.  Furthermore, the latest Belle  results \cite{Grygier:2017tzo}  infer  $R^{\nu\nu}_{K^{(\ast)}} <3.9 (2.7)$ at $90\%$ C.L.

\subsection{Semileptonic \texorpdfstring{$B$ Decays: $B\to D^{(\ast)}\ell\overline{\nu}$}{Semi}}
As already pointed out,  in addition to $R_2$ LQ, $S_3$ LQ can contribute to $R_D-R_{D^{(\ast)}}$. As argued, the corresponding contributions from $S_3$ are negligible. On the other hand, in this set-up, $S_3$ LQ can potentially contribute to $B\to D^{(\ast)}\ell\overline{\nu}$, with $\ell=e, \mu$. Belle Collaboration measured these lepton universality ratios in $e$ and $\mu$, and their results are given by: $R^{\mu/e}_{D^{(\ast)}}=0.995\pm 0.022\pm 0.039$ and 
$R^{e/\mu}_{D^{(\ast)}}=1.04\pm 0.05\pm 0.01$.  Both these measurements are well consistent with unity. In the current framework, these lepton universality ratios receive additional contributions that are given as: 
\begin{align}
\frac{R^{\mu/e}_{D^{(\ast)}}}{R^{\mu/e}_{D^{(\ast)},SM}}=\frac{1-Re\left(y_{b\mu}y^{\prime}_{c\mu}\right)}{1-Re\left(y_{be}y^{\prime}_{ce}\right)};\;\;\;\;
y^{\prime}_{q\ell}=(V^{\ast}y)^{\ast}_{q\ell}\frac{v^2}{2V_{cb}m^2_{S_3}}.
\end{align}

\subsection{Semileptonic \texorpdfstring{$K$ Decays: $K\to \ell\overline{\nu}$}{SemiKell}}
Lepton flavor universality in  kaon decays has also be measured with great accuracy. One such relevant observable in our scenario is the following ratio \cite{Fajfer:2015ycq}:
\begin{align}
R^{K}_{e/\mu}=\frac{\Gamma(K^-\to e^-\overline{\nu})}{\Gamma(K^-\to \mu^-\overline{\nu})}.    
\end{align}
The SM prediction and the 
experimental measured values of this ratios are given as follows \cite{Cirigliano:2007xi}:
\begin{align}
R^{K(exp)}_{e/\mu}=(2.477\pm 0.001)\times 10^{-5}, R^{K(exp)}_{e/\mu}=(2.488\pm 0.010)\times 10^{-5}.    
\end{align}
For our NP scenario, this constraint  can be recast into an upper  bound on the associated Yukawa coupling \cite{Fajfer:2015ycq, Dorsner:2017ufx}: $|y_{b\mu}|\lesssim 0.5 (m_S/TeV)$. This bound also plays a crucial role, since it is directly correlated with the explanation of $R_{K^{(\ast)}}$ ratios.

\subsection{Pseudoscalar Meson Decays:  \texorpdfstring{$P\to\ell^-\ell^{\prime+}$}{PseudoM}}
NP with  scalar LQs can have stringent constraints on Yukawa couplings as a result of leptonic decays of pseudoscalar mesons. The decay width associated with the process $P \to \ell^- \ell^{\prime+}$ can be written as \cite{Becirevic:2016zri}:
\begin{align}
\Gamma_{P \to \ell^{-}\ell^{\prime+}} &= 
f_{P}^2 m_{P}^3 \frac{G_F^2 \,\alpha_{e}^2}{64\pi^3}  
\left|V_{qj}V_{qi}^*\right|^2
f(m_{P},m_\ell,m_{\ell^{\prime}})
\scalemath{0.8}{
\left(
\left|
\frac{(m_\ell-m_{\ell^{\prime}})}{m_{P}} 
\left(C^{ij;\ell \ell^{\prime}}_{9}\right)
\right|^2 +\left| 
\frac{(m_\ell+m_{\ell^{\prime}})}{m_{P}} 
\left(C^{ij;\ell \ell^{\prime}}_{10}\right)
\right|^2 
\right)
},
\end{align}
here, $q$ refers to up-type quarks
and the function $f$ is defined as: 
\begin{align}
f(m_{P},m_\ell,m_{\ell^{\prime}}) = 
\sqrt{[1-(m_\ell-m_{\ell^{\prime}})^2/m_{P}^2] 
[1-(m_\ell+m_{\ell^{\prime}})^2/m_{P}^2)]}.
\end{align}
Whereas the leptonic decays $B_s^0 \to \ell^\pm \ell^\mp$  are well predicted in the SM, at present, only the $B_s \to \mu^+ \mu^-$ decay mode has been observed, which is in good agreement with the   SM prediction \cite{Bobeth:2013uxa}: $Br(B_s \to \mu^+ \mu^-)_{SM}=(3.65\pm 0.23)\times 10^{-9}$.  The observed value for the corresponding branching fraction is $Br(B_s \to \mu^+ \mu^-)_{exp}=(3.0\pm 0.6)\times 10^{-9}$, as reported by the LHC collaborations  \cite{Aaij:2017vad}. Furthermore, we also take into account the current limit of the lepton flavor violating $B$ meson decay with $e\mu$ final state \cite{Tanabashi:2018oca}: $Br(B_s\to \mu^{\pm}e^{\mp})< 5.4\times 10^{-9}$.

\subsection{Muon Anomalous Magnetic Moment: \texorpdfstring{$(g-2)_{\mu}$}{AMM}}
Anomalous magnetic moment (AMM) of the muon is one of the most preciously measured quantities in particle physics. Its value measured at the Brookhaven National Laboratory \cite{Bennett:2006fi} corresponds to a deviation of $3.7\sigma$ compared to the SM prediction. Their measured value is given as follows: 
\begin{align}
\Delta a_{\mu} =(2.74\pm 0.73) \times 10^{-9}.  \label{amm}  
\end{align}
Both $R_2$ and $S_3$ LQs can in principle contribute to the AMM of the muon, which we derive to be \cite{Cheung:2001ip, Dorsner:2016wpm}:
\begin{align}
\Delta a^{R_2}_{\mu}=-\frac{3m^2_{\mu}}{32 \pi^2} \frac{|y^L_{22}|^2}{m^2_{R^{5/3}}},\;\;\; 
\Delta a^{S_3}_{\mu}=-\frac{3m^2_{\mu}}{32 \pi^2} \frac{|y_{22}|^2+|y_{32}|^2}{m^2_S}.
\end{align}
Both of these provide wrong contributions to the muon AMM, hence the observed value quoted in Eq. \eqref{amm} cannot be incorporated within our set-up. Moreover, due to large masses of the LQs, this observation provides very weak bounds on the above Yukawa couplings.   

\subsection{Bounds from LHC Searches}  
In this section, we investigate the LHC constraints on the leptoquark (LQ) masses at Large Hadron Collider (LHC) for the relevance of B-decays in this model. At hadron colliders, LQ is pair produced via $gg$ or $q \bar{q}$ fusion. $q\bar{q}\to$ LQ LQ$^\dagger$ with a $t$-channel is highly suppressed compared to QCD induced production. Note, the coupling is assumed to be small in the production cross-section, $gg (q\bar{q})\to$ LQ LQ$^\dagger$, and it is a QCD driven process, entirely determined by the LQ mass and strong coupling constant. In both ATLAS and CMS, there are searches for the LQ pair production in different decay modes, LQ$^\dagger$ LQ $\to q \bar{q} \ell \bar{\ell}, q \bar{q} \nu \bar{\nu}$  associated with first \cite{Aaboud:2019jcc, Sirunyan:2018btu}, second \cite{Aaboud:2019jcc, Khachatryan:2015vaa,  Sirunyan:2018kzh, Sirunyan:2018ryt}, and third  \cite{Sirunyan:2018kzh, Sirunyan:2018vhk, Aaboud:2019bye} generations of quarks and leptons.    
\begin{table}[!t]
    \centering
\begin{tabular}{|c||c|c|c||c|}
\hline 
\textbf{Decays}  & \multicolumn{3}{|c||}{\textbf{Scalar LQ (GeV)}} & $\mathcal{L}_{\textbf{int }} $  \\
\cline{2-4}
& $\beta = 1$ & $\beta = 0.8$ & $\beta = 0.5$  & \\ \hline \hline
$j j \nu \bar{\nu}$ &  $ 980 $ & $810$ & $635$ & $35.9 \, \mathrm{fb}^{-1}$ \cite{Sirunyan:2018kzh} \\ \hline
$j j e \bar{e}$ & $1435$  & $1360$  & $1200$ &  $35.9 \, \mathrm{fb}^{-1}$ \cite{Sirunyan:2018btu}\\
$j j \mu \bar{\mu}$ & $1525$   & $1445$  & $1280$ &  $36.1 \, \mathrm{fb}^{-1}$ \cite{Sirunyan:2018ryt} \\ \hline
$t \bar{t} \tau \bar{\tau}$ & $930$   & $850$  & $730$ &  $36.1 \, \mathrm{fb}^{-1}$ \cite{Aaboud:2019bye}\\
$t \bar{t} \nu \bar{\nu}$ &  $1020$  & $940$  & $812$ &   $35.9 \, \mathrm{fb}^{-1}$ \cite{Sirunyan:2018kzh}\\
$b \bar{b}  \tau \bar{\tau}$ & $1025$  & $965$  & $835$ & $36.1 \, \mathrm{fb}^{-1}$ \cite{Aaboud:2019bye, Sirunyan:2018vhk} \\
\hline
\end{tabular}
     \caption{Summary of scalar LQ mass bound with different decay modes for branching ratio $\beta$ (1, 0.8, 0.5) at LHC. '$j$' denotes light quarks $u, d, c, s$. In the first column, we list the possible final states, whereas second, third, and fourth column shows the current limits on the masses of scalar LQs for three benchmark branching ratios. The last column represents the LHC luminosity for each search with the corresponding references.}
    \label{tab:LQlimit}
\end{table}

In Table \ref{tab:LQlimit}, we show the current limits on the masses of first, second, and third generations scalar LQs for three benchmark values of branching ratio $\beta=$ (1, 0.8, 0.5). For the relevance of $R_D-R_{D^{\ast}}$, we take a minimal choice of Yukawa coupling matrices given in Eq.~\ref{TX-I} and Eq.~\ref{TX-II}, and look at the corresponding decay modes. Thus, $R_2^{5/3}$ with Yukawa coupling $y^L_{ij}$ $(i=c, j = e, \mu, \tau)$ and $(Vy^R)_{33}$  lead to $jj\ell \bar{\ell}$ and $t\bar{t}\tau \bar{\tau}$ final states, respectively from the decay of the pair produced LQs. Similarly, $R_2^{2/3}$ with Yukawa coupling $(y^L U)_{ij}$ $(i=c, j = \nu_e, \nu_\mu, \nu_\tau)$ and $(y^R)_{33}$  lead to $jj\nu \bar{\nu}$ and $b\bar{b}\tau \bar{\tau}$ final states, respectively. It is worth mentioning that $S_3$ LQ, relevant for $R_K-R_{K^*}$, has all the above-mentioned decay modes. However, we take the mass of the $S_3$ to be 2.5 TeV, which automatically satisfies all current LHC bounds. 

To obtain the pair-produced  LQ constraints at large Yukawa couplings, we quote Ref. \cite{Schmaltz:2018nls} for various $j\ell$ and $b\ell$ type  LQs. Here, $t$-channel diagram interferes with $gg$ and $q\bar{q}$ fusion diagrams. The bounds indicated by Ref. \cite{Schmaltz:2018nls} shows that t-channel starts to dominate for a LQ that couples to first generation quarks with coupling of $y \gtrsim 1$. However, LQ coupling to heavier quarks has main contribution from $t$-channel only for $y \gg 1$. It is also to be pointed that Drell-Yan-like production with the $\ell^+ \ell^-$ final state puts bounds on LQ masses and the associated Yukawa couplings. Such as, LQ coupling to $ce$, $c\mu$, $c\tau$, and $b\tau$ has a bound on the Yukawa couplings of $1.8, 1.5, 2.3,$ and $2.3$, respectively for 1 TeV LQ mass. The constraints on the Yukawa couplings and the bounds on the LQ masses discussed here will be used in the next section for our numerical analysis.

\section{Results and Discussions}\label{SEC-06}
In this section, we present the numerical results of the model introduced in Sec. \ref{SEC-02}. As already aforementioned, in search of finding the minimal parameters of the theory, our detailed numerical analysis results in two separate textures that are summarized in Eqs. \eqref{TX-I} and  \eqref{TX-II}. It is important to emphasize that we are interested in finding viable solutions where all the neutrino observables and the $R_{D^{(\ast)}}$, $R_{K^{(\ast)}}$ flavor ratios are reproduced within their $1\sigma$ measured values. For this numerical analysis, we fix the LQ masses to be $m_R=1$ TeV and $m_S=2.5$ TeV.  For the masses of the up-type quarks  entering in the neutrino mass matrix, we take their values from the PDG \cite{Esteban:2018azc}:  $m_u=2.16$ MeV,  $m_c=1.27$ GeV,  $m_t=172.9$ GeV, and consider the following values of the CKM entries:
\begin{align}
V= \left(
\begin{array}{ccc}
 0.974724 & 0.224837 & 0.00358919 \\
 -0.224837 & 0.974724 & 0.0416293 \\
 0.00854386 & -0.0416293 & 1 \\
\end{array}
\right).    
\end{align}
We remind the readers that the overall factor $m_0$ in front of the neutrino mass matrix Eq.~\eqref{numat} is a free parameter of the theory, which is required to be tiny $m_0\sim 10^{-8}$ GeV to satisfy the neutrino oscillation data. Then Eq. \eqref{numat} provides $\mu \lambda\sim 0.7$ GeV, which clearly shows that for the phenomenological analysis, the mixing among the scalars can be ignored entirely within our scenario. It is because the corresponding mixings are parametrized by the cubic coupling $\mu$ and the quartic coupling $\lambda$ present in the scalar potential.

As discussed in Sec.~\ref{SEC-03},  the solution to the $R_D-R_{D^{\ast}}$ observables in scenarios with significant contributions from   Wilson coefficient $C^{\tau}_L$ must be complex, which is always the case in our model. Consequently, we choose the parameter $y^R_{33}$ to be purely imaginary, and the rest of the parameters are all taken to be real. A fit to the $R_K-R_{K^{\ast}}$ observables within $1\sigma$ measured values demands  $y_{b\mu}y_{s\mu}$ to lie in between $[4.49\times 10^{-3}, 6.09\times 10^{-3}]$.  In this search strategy, we appropriately include all the relevant constraints listed in Sec. \ref{SEC-05} coming from various experimental results.  It is beyond the scope of this work to explore the entire parameter space; instead we employ an optimization technique that helps us efficiently explore different Yukawa textures and verify their viabilities as discussed below. 

Our numerical method is based on a constrained minimization where five of the neutrino observables ($\Delta m^2_{21}, \Delta m^2_{31}, \sin^2 \theta_{12}, \sin^2 \theta_{23}, \sin^2 \theta_{13}$), along with the four flavor ratios ($R_D, R_{D^{\ast}}, R_K, R_{K^{\ast}}$), are forced to stay close to their experimental measured values. Deviations from these values are parametrized by pulls defined by $P_i=(T_i - O_i)/E_i$, where $T_i$ is the theory prediction, $O_i$ is the experimental central value, and $E_i$ is the associated $1\sigma$ error corresponding to an observable $i$. A $\chi^2$-function is then formed out of these pulls $\chi^2[x_k]= \sum_i P_i[x_k]$ that enters the minimization. The theory parameters $x_k$ are further constrained by the other experimental measurements discussed in the previous section. Utilizing the set of formulas derived in Sec. \ref{SEC-05}, we impose those conditions to restrict the theory parameters $x_k$, such as $Br(\mu \to e \gamma)[x_k] < 4.2\times 10^{-13}$. By performing this constrained minimization for several different textures of Yukawa coupling matrices, we filter out scenarios that fail to give a reasonable fit. 

Corresponding to Eqs. \eqref{TX-I} and \eqref{TX-II}, the minimal and viable Yukawa textures that resulted from our numerical search, we provide one benchmark point for each in Eqs.  \eqref{BM-TX-I}  and \eqref{BM-TX-II}, respectively. A list of observables associated with these parameter sets is tabulated in Table \ref{result}.

\begin{align} 
&\texttt{BM-TX-I}:\label{BM-TX-I}  \\ &
y^R=\left(\begin{array}{ccc}
0 & 0 & 0 \\
0 & 0 & 0 \\
0 & 0 & 1.09527\;i
\end{array}\right),\;\;
y^L = \left(\begin{array}{ccc}
0 & 0 & 0 \\
4.0503\times 10^{-3} & -2.1393\times 10^{-2} & 1.2243 \\
0 & 0 & -4.9097\times 10^{-4}
\end{array}\right),\nonumber\\
&y = \left(\begin{array}{ccc}
0 & 0 & 0 \\
-4.5241\times 10^{-4}& -7.7187\times 10^{-3} & -4.6354\times 10^{-4} \\
0 & 6.8578\times 10^{-1} & 0
\end{array}\right),\;\; m_0=1.297\times 10^{-8}.\nonumber
\end{align}
\begin{align} 
&\texttt{BM-TX-II}:\label{BM-TX-II}\\&
y^R=\left(\begin{array}{ccc}
0 & 0 & 0 \\
0 & 0 & 0 \\
0 & 0 & 1.0945\;i
\end{array}\right),\;\;\;
y^L = \left(\begin{array}{ccc}
0 & 0 & 0 \\
-4.0823\times 10^{-3} & 2.3482\times 10^{-2} & 1.2250 \\
0 & 0 & -1.6522\times 10^{-2}
\end{array}\right),\nonumber\\
&y = -\left(\begin{array}{ccc}
0 & 0 & 0 \\
0 & 9.6924\times 10^{-2} & 1.2668\times 10^{-3} \\
7.2254\times 10^{-4} & 5.4592\times 10^{-2} & 0
\end{array}\right),\;\;\; m_0=4.409\times 10^{-9}.\nonumber
\end{align}
The $\hat{y}^L=y^LU$ matrices associated to these benchmark points are listed in Appendix \ref{A}.

\FloatBarrier
\begin{table}[b!]
\centering
\footnotesize
\resizebox{0.5\textwidth}{!}{
\begin{tabular}{|c|c|c|}
\hline
\textbf{Observables} & \texttt{BM-TX-I}& \textbf{{\tt BM-TX-II}}  \\ [1ex] \hline \hline
$\Delta m^2_{21} (eV^2)$&$7.39\times 10^{-5}$&$7.40\times 10^{-5}$ \\ \hline
$\Delta m^2_{31} (eV^2)$&$2.52\times 10^{-3}$&$2.52\times 10^{-3}$  \\ \hline
$\theta_{12}$&$33.86^{\circ}$& $33.87^{\circ}$ \\ \hline
$\theta_{23}$&$47.95^{\circ}$& $49.95^{\circ}$  \\ \hline
$\theta_{13}$&$8.61^{\circ}$&$8.61^{\circ}$    \\ \hline \hline
$R_D$&0.3485&0.3484  \\ \hline
$R_{D^{\ast}}$&0.2875&0.2875   \\ \hline
$C_9=-C_{10}$&-0.529&-0.530 \\ \hline \hline
$CR(\mu \to e)$&$3.31\times 10^{-15}$&$5.90\times 10^{-17}$    \\ \hline
$Br(\mu \to e\gamma)$&$1.34\times 10^{-14}$&$1.45\times 10^{-14}$  \\ \hline
$Br(\tau \to e\gamma)$&$4.11\times 10^{-10}$&$1.49\times 10^{-11}$  \\ \hline
$Br(\tau \to \mu\gamma)$&$1.47\times 10^{-11}$&$4.95\times 10^{-10}$   \\ \hline \hline
\end{tabular}
}
\caption{Fit values of some of the observables for benchmark points given in Eqs. \eqref{BM-TX-I} and  \eqref{BM-TX-II}.}
\label{result}
\end{table}

\begin{figure}[th!]
\centering
\includegraphics[scale=0.42]{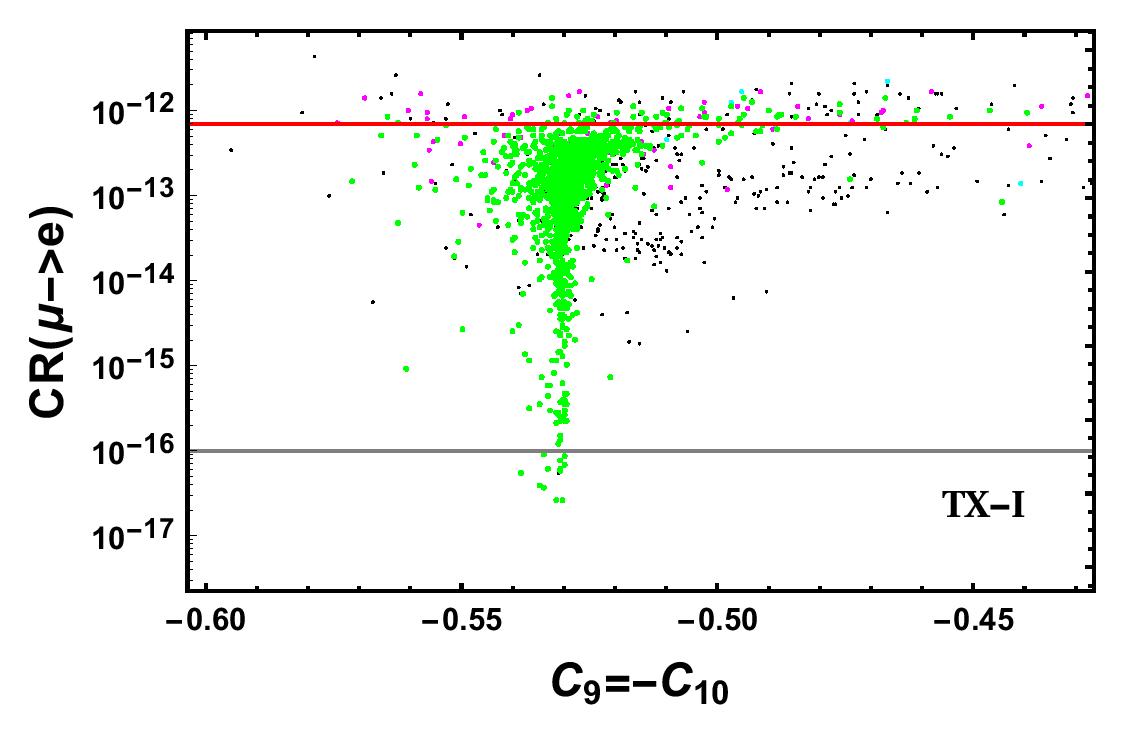}
\includegraphics[scale=0.42]{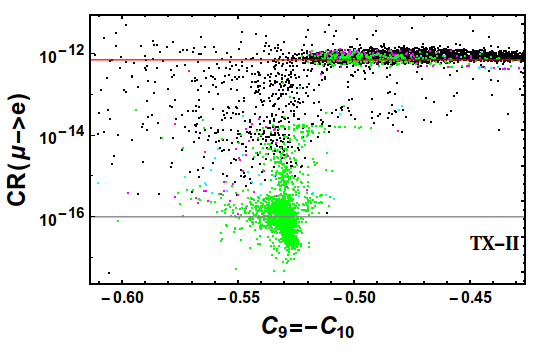}
\caption{Biased scattered plot resulting from our constrained minimization procedure in the $C_9 - CR(\mu\to e)$ plane.  The solid red line corresponds to the present bound of  $CR(\mu\to e)<7\times 10^{-13}$ from SINDRUM II \cite{Bertl:2006up}, and the gray solid line is for the most stringent bound of $CR(\mu\to e)< 10^{-16}$ from future projected sensitivity \cite{Kurup:2011zza, Cui:2009zz, Chang:2000ac, Adamov:2018vin, Bartoszek:2014mya, Pezzullo:2018fzp, Bonventre:2019grv}. Green dots correspond to solutions where all the five neutrino observables are reproduced within their $1\sigma$ measured values. Similarly magenta (cyan) dots represent solutions that are reproduced within their  $2\sigma$ ($3\sigma$) experimental values. Solutions that do not return neutrino mass squared differences and mixing angles within $3\sigma$ values are represented with black dots. }\label{plot1}
\end{figure}

\begin{figure}[th!]
\centering
\includegraphics[scale=0.42]{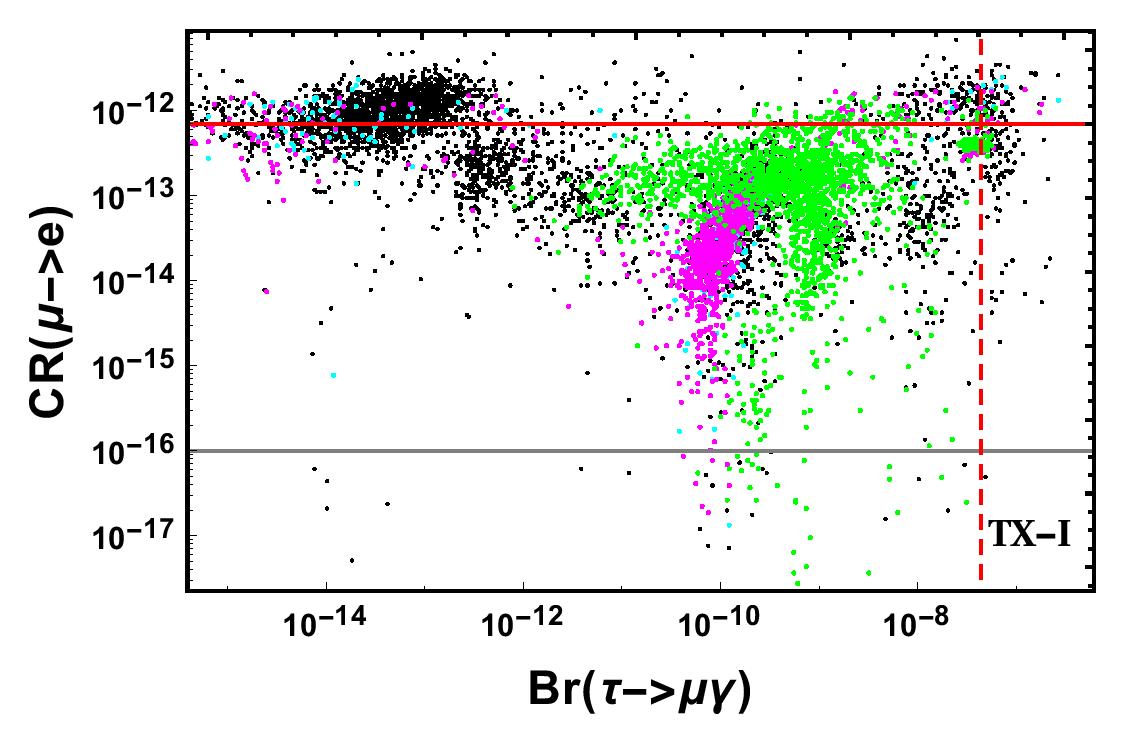}
\includegraphics[scale=0.42]{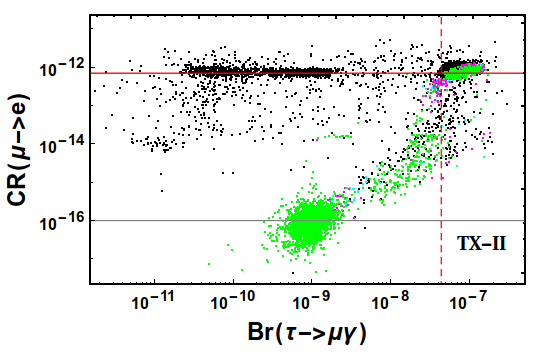}
\caption{Biased scattered plot in the $Br(\tau\to \mu \gamma)- CR(\mu\to e)$ plane, for details see text. The solid red horizontal line corresponds to the present bound of  $CR(\mu\to e)<7\times 10^{-13}$ from SINDRUM II \cite{Bertl:2006up}, and the gray solid horizontal line is for the most stringent bound of $CR(\mu\to e)< 10^{-16}$ from future projected sensitivity \cite{Kurup:2011zza, Cui:2009zz, Chang:2000ac, Adamov:2018vin, Bartoszek:2014mya, Pezzullo:2018fzp, Bonventre:2019grv}. The vertical red dashed line corresponds to present experimental limit  $Br(\tau \to \mu \gamma) < 4.4\times 10^{-8}$ \cite{Aubert:2009ag}. The color code is the same as that of Fig. \ref{plot1}.}\label{plot4}
\end{figure}

\begin{figure}[th!]
\centering
\includegraphics[scale=0.42]{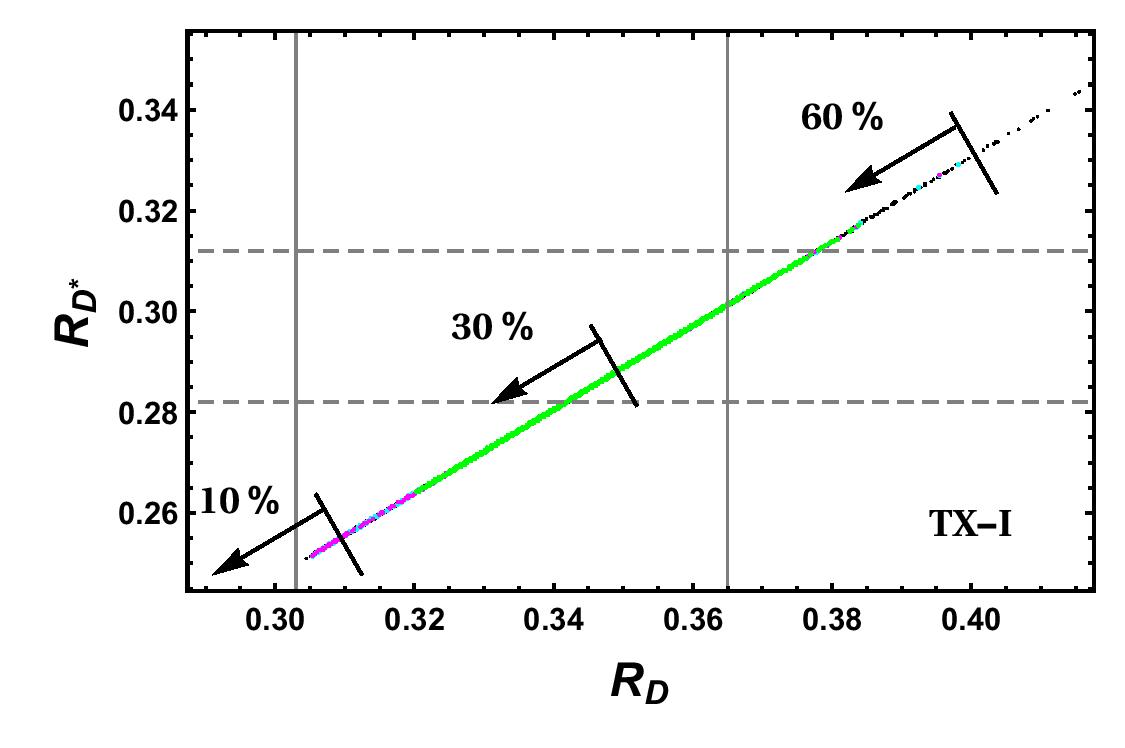}
\includegraphics[scale=0.42]{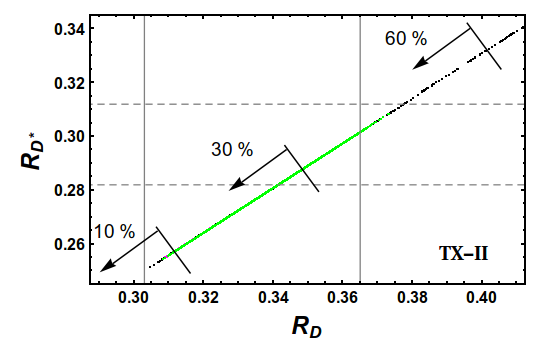}
\caption{Biased scattered plot in the $R_D-R_{D^{\ast}}$ plane. The region in between the vertical (horizontal) solid gray (dashed gray) lines correspond to $1\sigma$ allowed range for $R_D$ ($R_{D^{\ast}}$) ratio.   We have also indicated three different branching ratios of $Br(B_c\to \tau\nu)$ corresponding to $10\%, 30\%$, and $60\%$ within the present set-up.  The color code is the same as that of Fig. \ref{plot1}.}\label{plot3}
\end{figure}

For each of these Yukawa textures, we iterate over the aforementioned constrained minimization procedure ten thousand times to explore the correlation among different physical quantities. These results are presented in Figs. \ref{plot1}, \ref{plot2}, \ref{plot3}, \ref{plot4}, \ref{plot5}, and \ref{plot6}. It must be understood that these plots do not correspond to a random scan over all the theory parameters, rather a biased scattered plot as a result of our constrained optimization procedure. Due to strong correlations among many different observables, 
random scans most likely return parameters that fail to provide good fits to all observables at once. On the contrary,  constrained minimization is more likely to force towards the valid region in the parameter space.  In these plots, green dots correspond to solutions where all the five neutrino observables are reproduced within their $1\sigma$ measured values. Similarly, magenta (cyan) dots represent solutions that are reproduced within their  $2\sigma$ ($3\sigma$) experimental values. Solutions that do not return neutrino mass squared differences and three mixing angles within their $3\sigma$ measured values are represented with black dots. These correlations are obtained since neutrino masses and mixings, flavor observable ratios $R_{D^{(\ast)}}$, $R_{K^{(\ast)}}$, and other flavor violating processes are all intertwined within our framework. 


As can be seen from these plots, experimental results of charged lepton flavor violating processes namely $\mu\to e$ conversion in nuclei, and $\ell\to\ell^{\prime}\gamma$ decays have strong implications in our scenario.   Future improved sensitivity of these experiments will probe most part of the parameter space. For example, Fig. \ref{plot1} demonstrates the correlation of $R_{K^{(\ast)}}$ ratios with $\mu\to e$ in gold. This shows that for both the textures, specially for {\tt TX-I}, only a small parameter space will remain viable, since $\mu\to e$ conversion rate cannot be made arbitrarily small and still be consistent with observed $R_{K^{\ast}}$ ratio. Strong correlations of different lepton flavor violating processes that are typical within our scenario are depicted in Figs. \ref{plot2}, \ref{plot4}, \ref{plot5}, and \ref{plot6}.  Furthermore, from Fig. \ref{plot3}, it can be inferred that $Br(B_c\to \tau \nu)> 10\%$ must be realized to get an acceptable solution to $R_{D^{(\ast)}}$.

For each set of Yukawa structures, as shown in Eq.~\eqref{BM-TX-I} and Eq.~\eqref{BM-TX-II}, we set the limits on the masses of the leptoquarks. $R_2^{5/3}$ has decay modes to $j j \tau \bar{\tau}$ and $t \bar{t} \tau \bar{\tau}$ with corresponding branching ratios $\beta=$  0.56 and 0.44, respectively. Similarly, $R_2^{2/3}$ primarily decays to $jj \nu \bar{\nu}$ with $\beta$ of 0.55 and $b\bar{b} \tau \bar{\tau}$ with $\beta$ of 0.45. With these given decay modes along with their associated branching fractions, the most robust bound on LQ mass is 810 GeV that comes from pair produced $R_2^{2/3}$ LQs decaying to $b\bar{b} \tau \bar{\tau}$. However, we have set the LQ mass for $R_2$ at 1 TeV for demonstration, which automatically satisfies the mass limit from LHC, as quoted in Table \ref{tab:LQlimit}. Since, LQs in both of these textures only decay to heavier quarks (c, b, t), LQ pair production $t$-channel diagram and Drell-Yan-like production with $\ell^+ \ell^-$ final state do not put important constraints with the choice of Yukawa couplings presented in Eqs.~\eqref{BM-TX-I} and \eqref{BM-TX-II}. From the above analysis it can be inferred that consistent explanation of $R_{D^{\ast}}$ ratios requires the mass of the $R_2$ LQ in the range $\sim$ 800-1000 GeV without making the corresponding Yukawa couplings much larger than unity. A scalar LQ of this type can be directly probed at the LHC by the upcoming  high luminosity run. A detailed collider study is beyond the scope of this work.

Here we comment on the effects of the $\mathcal{O}(1)$ couplings required for $R_{D^{(*)}}$ explanations. In certain scenarios large Yukawa couplings to the third generation lepton (tau) can induce universal contributions to $C^{\ell\ell}_{9,10}$ for light lepton generations ($\ell=e, \mu$) via the one-loop off-shell photon penguin diagram \cite{Bobeth:2014rda} ($Z$-boson penguin diagrams can be safely neglected). Which under renormalization group equation running down to the B-meson mass scale can give significant contribution to the $b\to s\mu\mu$ transition. This plays a vital role in scenarios with vector LQ where the same couplings that explain the $R_{D^{(*)}}$ induce such one-loop photon penguin diagram \cite{Crivellin:2018yvo}. Concerning the scalar LQs, similar logarithmic enhanced contributions can also be realized for $S_3$ LQ \cite{Crivellin:2019dwb}. However, in our scenario with $R_2$ ($S_3$) LQ, no such one-loop  photon penguin diagram can be drawn due to our choice of $y^R_{s\tau}=0$ ($y_{b\tau}=0$).

\section{Conclusion}\label{SEC-07}
In conclusion, we have proposed an economic model where three of the problems are interconnected and have a common solution. The issue of generating neutrino masses is directly linked with the persistent observation of $B$ meson decay anomalies in the $R_D, R_{D^{\ast}}$ and $R_K, R_{K^{\ast}}$ ratios over the last several years. Our proposed model consists of three beyond the SM scalar color triplets; they are a singlet $\chi_1$, a doublet $R_2$, and a triplet $S_3$ under the  $SU(2)_L$.  $R_{D^{(\ast)}}$ and $R_{K^{(\ast)}}$ anomalies are accommodated by $R_2$ and  $S_3$ scalar leptoquarks, respectively, whereas neutrinos receive tiny masses via quantum corrections where all these beyond SM scalars are running inside the loops. By properly taking into account all the relevant experimental constraints, we have demonstrated the viability of our scenario that is achieved with only a limited number of Yukawa parameters. This is a highly non-trivial task since the same parameters that explain neutrino oscillation data, and $R_{D^{(\ast)}}, R_{K^{(\ast)}}$ flavor ratios, also lead to many other flavor violating processes that are strongly constrained by experimental observations.  Low energy experiments that are searching for rare lepton flavor violating decays provide tight constraints, and future improved sensitivity of these measurements will probe substantial part of the parameter space of the model. For the explanation of $R_{D^{(\ast)}}$, since complex Yukawa coupling of order unity is required, such a scenario can be directly probed in the future experiments by precise measurements of electric dipole moments \cite{Dekens:2018bci}. In our set-up, the explanations of both $R_D$ and $R_{D^{\ast}}$ ratios within their $1\sigma$ allowed range require the branching ratio  $Br(B_c\to \tau\nu)$ much larger than $10\%$, which in principle can be ruled out in future by reducing the uncertainties associated with QCD calculations. Furthermore, our description of flavor anomalies, particularly for $R_{D^{(\ast)}}$ ratios predict LQ states around the TeV scale that are directly accessible at the LHC. 

\section*{Acknowledgments}
We thank Ahmed Ismail and K. S. Babu for useful discussion. The work of A.T. was supported in part by the US Department of Energy Grant Number DE-SC 0016013. 

\begin{appendices}
\renewcommand\thesection{\arabic{section}}
\renewcommand\thesubsection{\thesection.\arabic{subsection}}
\renewcommand\thefigure{\arabic{figure}}
\renewcommand\thetable{\arabic{table}}
\section{}\label{A}
In this appendix we present the numerical values of the $\hat{y}^L$ matrices for the two benchmark points considered in the main text.  
\begin{align}
&\texttt{BM-TX-I}:\;\;\;
y^L U=\begin{pmatrix}
0&0&0\\
-0.411367&0.836255&0.79438\\
1.6229\times 10^{-4}&-3.3020\times 10^{-4}&-3.2508 \times 10^{-4}
\end{pmatrix}.
\\
&\texttt{BM-TX-II}:\;\;\;
y^L U=\begin{pmatrix}
0&0&0\\
-0.397435&0.810018&0.8289928\\
5.4623\times 10^{-3}&-1.1111\times 10^{-2}&-1.0939 \times 10^{-2}
\end{pmatrix}.
\end{align}

\clearpage
\section{}\label{B}
\begin{figure}[th!]
\centering
\includegraphics[scale=0.42]{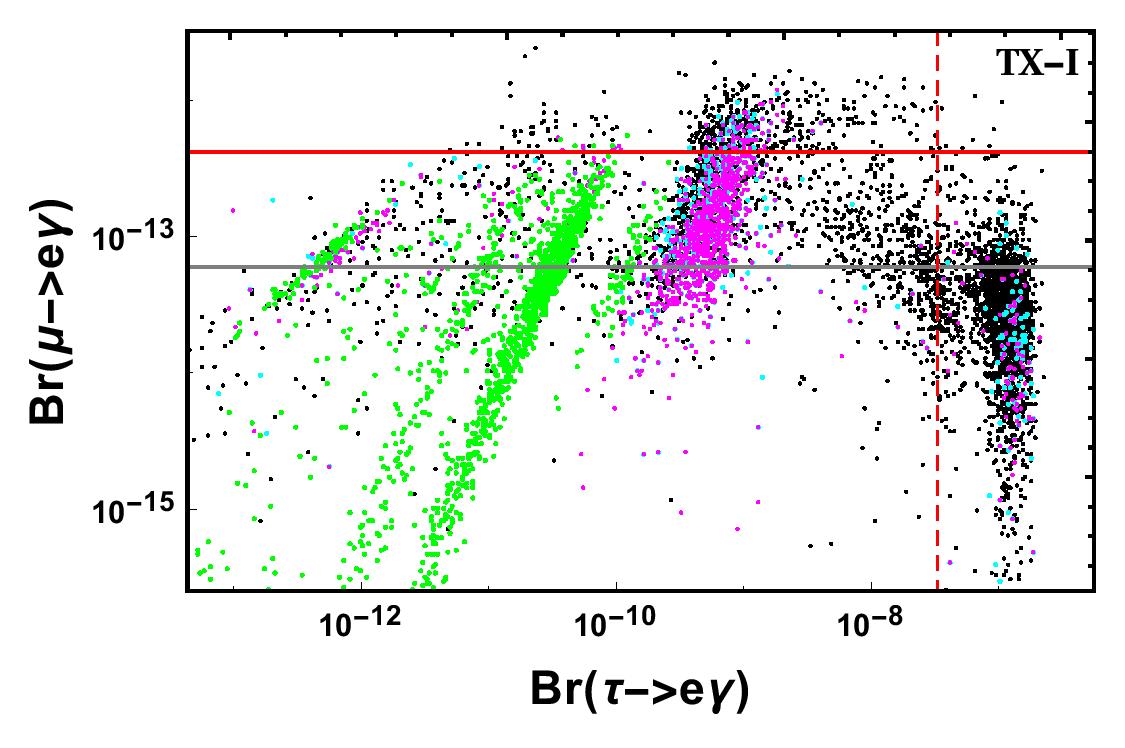}
\includegraphics[scale=0.42]{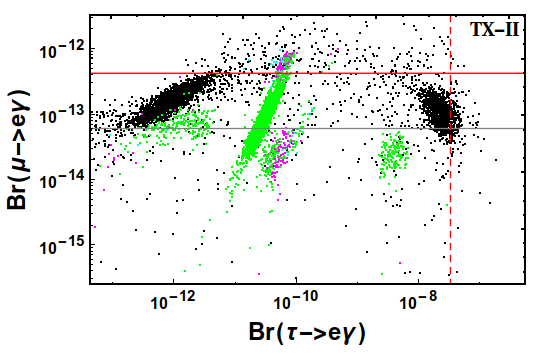}
\caption{Biased scattered plot in the $Br(\tau\to e \gamma)- Br(\mu\to e \gamma)$ plane, for details see text. The red (gray) solid horizontal line represents current (future projected sensitivity) bound  $Br(\mu \to e \gamma) < 4.2\times 10^{-13}$ ($6\times 10^{-14}$) \cite{TheMEG:2016wtm}. The vertical red dashed line corresponds to present experimental limit  $Br(\tau \to e \gamma) < 3.3\times 10^{-8}$ \cite{Aubert:2009ag}. The color code is the same as that of Fig. \ref{plot1}.}\label{plot5}
\end{figure}

\begin{figure}[th!]
\centering
\includegraphics[scale=0.42]{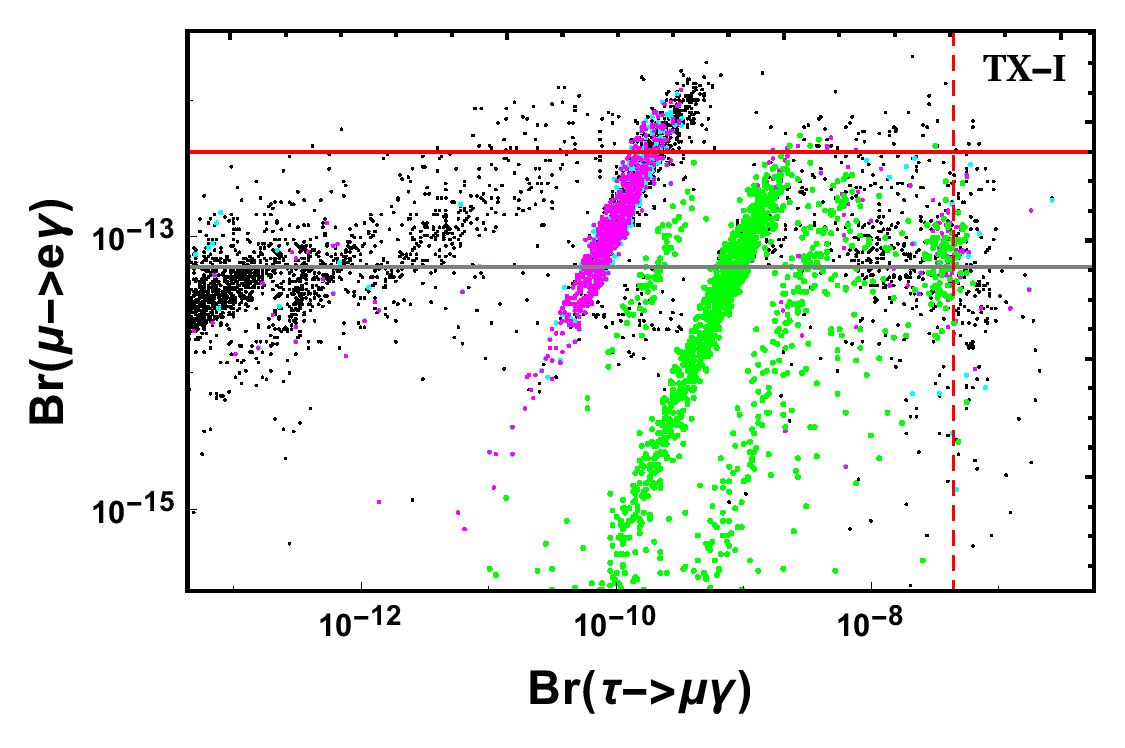}
\includegraphics[scale=0.42]{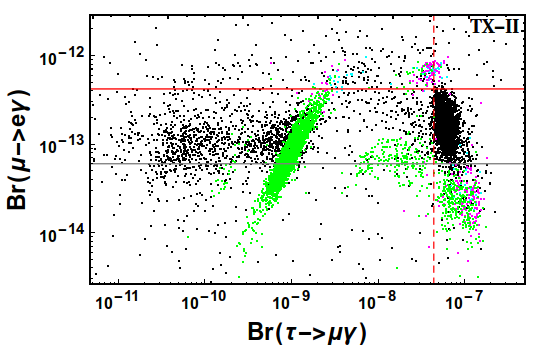}
\caption{Biased scattered plot in the $Br(\tau\to \mu \gamma)- Br(\mu\to e \gamma)$ plane, for details see text. The red (gray) solid horizontal line represents current (future projected sensitivity) bound  $Br(\mu \to e \gamma) < 4.2\times 10^{-13}$ ($6\times 10^{-14}$) \cite{TheMEG:2016wtm}. The vertical red dashed line corresponds to present experimental limit  $Br(\tau \to \mu \gamma) < 4.4\times 10^{-8}$ \cite{Aubert:2009ag}. The color code is the same as that of Fig. \ref{plot1}.}\label{plot2}
\end{figure}

\begin{figure}[t!]
\centering
\includegraphics[scale=0.42]{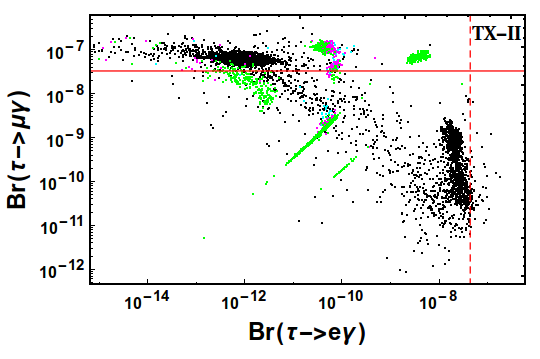}
\includegraphics[scale=0.42]{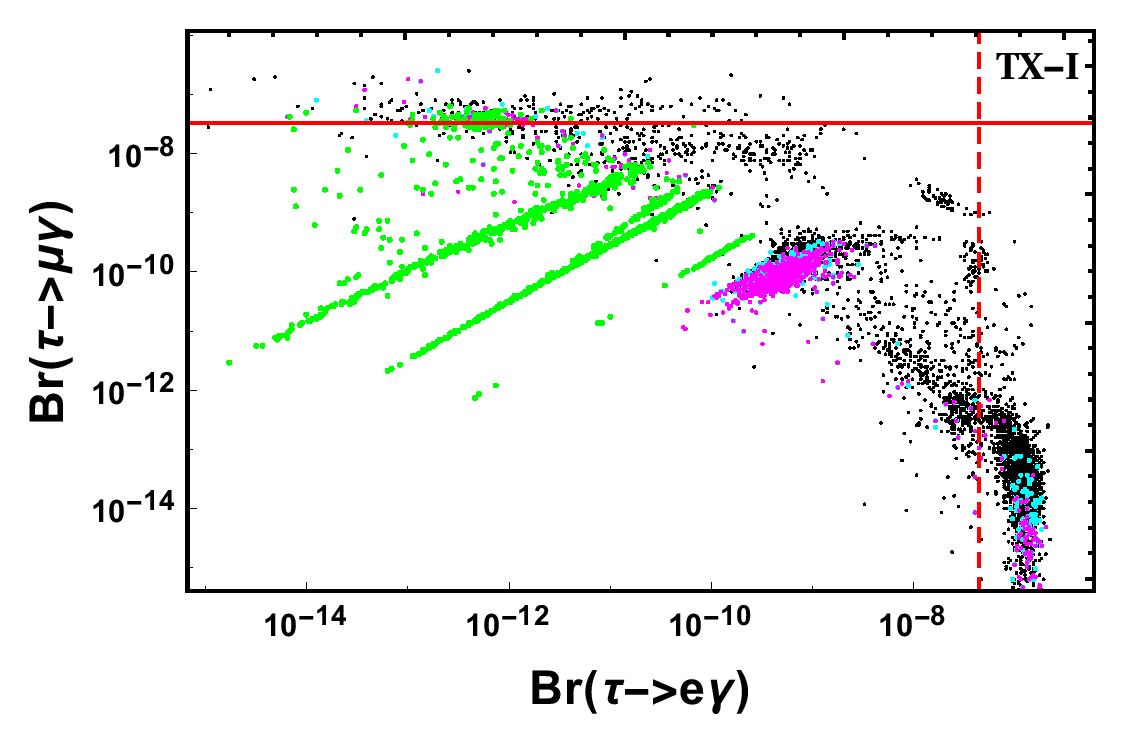}
\caption{Biased scattered plot in the $Br(\tau\to e \gamma)- Br(\tau\to \mu \gamma)$ plane, for details see text. The red solid horizontal line represents current  bound  $Br(\tau \to \mu \gamma) < 4.4\times 10^{-8}$ and the vertical red dashed line corresponds to present experimental limit  $Br(\tau \to \mu \gamma) < 3.3\times 10^{-8}$ \cite{Aubert:2009ag}. The color code is the same as that of Fig. \ref{plot1}.}\label{plot6}
\end{figure}

Correlations among few different lepton flavor violating processes are presented in Figs. \ref{plot5}, \ref{plot2}, and \ref{plot6}.

\end{appendices}

\vspace{5cm}
\bibliographystyle{utphys}
\bibliography{references}
\end{document}